\documentclass[a4paper,11pt]{article}
\usepackage{amsmath, amsthm, amssymb}
\usepackage{bm}
\usepackage{graphicx}
\usepackage{fullpage}
\usepackage[colorlinks,linkcolor=blue,citecolor=blue,urlcolor=magenta,linktocpage,plainpages=false]{hyperref}\usepackage{color}
\usepackage{xspace}
\usepackage{algorithm}
\usepackage{url}
\usepackage{algpseudocode}
\usepackage{multicol}
\usepackage{caption}
\usepackage[normalem]{ulem}
\usepackage{thmtools,thm-restate}

\usepackage[nameinlink]{cleveref}
\usepackage{comment}

\newtheorem{lemma}{Lemma}
\newtheorem{thm}{Theorem}
\newtheorem{cor}{Corollary}
\newtheorem{prop}{Proposition}

\newtheorem{claim}{Claim}
\newtheorem{definition}{Definition}
\newtheorem{obs}{Observation}
\newtheorem{example}{Example}

\newcommand{\E}[0]{\ensuremath \mathbb{E}}
\newcommand{\R}[0]{\ensuremath \mathbb{R}}
\newcommand{\e}{\varepsilon}

\newcommand{\ones}{\bm{1}}
\newcommand{\blue}[1]{\textcolor{blue}{#1}}
\newcommand{\red}[1]{{\color{red} #1}}
\newcommand{\cL}{\mathcal{L}}

\newcommand{\mom}[1]{{\left\vert\kern-0.25ex\left\vert\kern-0.25ex\left\vert #1 \right\vert\kern-0.25ex\right\vert\kern-0.25ex\right\vert}}

\usepackage{tikz}
\usetikzlibrary{arrows,positioning}
\usepackage[shortlabels]{enumitem}
\usetikzlibrary{patterns}
\usetikzlibrary{decorations.pathreplacing}
\usetikzlibrary{calc}
\usepackage[outline]{contour}
\contourlength{1.2pt}

\newcounter{mynotes}
\newcommand{\knote}[1]{\addtocounter{mynotes}{1}{{\bf !}}%
\marginpar{\scriptsize  {\arabic{mynotes}.\ {\sf \textcolor{red}{#1}}}}}

\newcommand{\rnote}[1]{\addtocounter{mynotes}{1}{{\bf !}}%
\marginpar{\scriptsize  {\arabic{mynotes}.\ {\sf \textcolor{blue}{#1}}}}}

\newcommand{\mnote}[1]{\addtocounter{mynotes}{1}{{\bf !}}%
\marginpar{\scriptsize  {\arabic{mynotes}.\ {\sf \textcolor{magenta}{#1}}}}}

    \renewcommand{\knote}[1]{}
    \renewcommand{\rnote}[1]{}
    \renewcommand{\mnote}[1]{}

    \newcommand{\replace}[2]{#1}

\newcommand{\C}{\mathcal{C}}

\newcommand{\imu}{\mu^{-1}}

\newcommand{\alg}{\textup{\sf Alg}\xspace}
\newcommand{\opt}{\textup{\sf OPT}\xspace}
\newcommand{\OPT}{\opt}
\newcommand{\lpconfigmin}{\ensuremath{\textup{\sf LP}_{mach}}}

\newcommand{\problem}{{\sc Failover}\xspace}
\newcommand{\minoff}{\textsc{OffMinFailover}\xspace}

\newcommand{\matchalg}{\textup{\sf MatchConfigs}\xspace}
\newcommand{\oneround}{\textup{\sf OneRound}\xspace}
\newcommand{\offlineminmach}{\textup{\sf OffMinFailoverAlg}\xspace}
\newcommand{\minoffalg}{\offlineminmach}
\newcommand{\failoverstochastic}{\textup{\sf FailoverStochastic}\xspace}

\newcommand{\mach}{\ensuremath{\opt_{mach}}}

\newcommand{\cR}{\mathcal{R}}
\newcommand{\size}{size}

\newcommand{\mubar}[1]{\underbar{$#1$}}

\newcommand{\cst}{cst}
\newcommand{\csta}{cst_1}
\newcommand{\cstc}{cst_2}
\newcommand{\cstd}{cst_3}
\newcommand{\cstf}{cst_4}
\newcommand{\cstg}{cst_5}

\newcommand{\N}{\mathbb{N}}
\newcommand{\cdf}{\textup{cdf}}

\def\hasmain{}

\title{Online Demand Scheduling with Failovers}
\date{\today}
\author{Konstantina Mellou \\ Microsoft Research Redmond \\ kmellou@microsoft.com
	\and Marco Molinaro \thanks{Work done while visiting Microsoft Research Redmond} \\ PUC-Rio \\ molinaro.marco@gmail.com
	\and Rudy Zhou \footnotemark[1] \\ Carnegie Mellon \\ rbz@andrew.cmu.edu}
\begin{document}
\maketitle

\begin{abstract}
	
	
	Motivated by cloud computing applications, we study the problem of how to optimally deploy new hardware subject to both power and robustness constraints. To model the situation observed in large-scale data centers, we introduce the \emph{Online Demand Scheduling with Failover} problem. There are $m$ identical devices with capacity constraints. Demands come one-by-one and, to be robust against a device failure, need to be assigned to a pair of devices. When a device fails (in a failover scenario), each demand assigned to it is rerouted to its paired device (which may now run at increased capacity). The goal is to assign demands to the devices to maximize the total utilization subject to both the normal capacity constraints as well as these novel failover constraints. These latter constraints introduce new decision tradeoffs not present in classic  assignment problems such as the Multiple Knapsack problem and AdWords.

    
    
    In the worst-case model, we design a deterministic  $\approx \frac{1}{2}$-competitive algorithm, and show this is essentially tight. To circumvent this constant-factor loss, which in the context of big cloud providers represents substantial capital losses, we consider the stochastic arrival model, where all demands come i.i.d. from an unknown distribution. In this model we design an algorithm that achieves a sub-linear additive regret (i.e. as $\opt$ or $m$ increases, the multiplicative competitive ratio goes to $1$). This requires a combination of different techniques, including a configuration LP with a non-trivial post-processing step and an online monotone matching procedure introduced by Rhee and Talagrand.     
    
\end{abstract}

\thispagestyle{empty}
\clearpage

\setcounter{tocdepth}{2}
{\small
   \tableofcontents
} 

\thispagestyle{empty}

\clearpage

\setcounter{page}{1}







\section{Introduction}


A critical challenge faced by cloud providers is how to optimally deploy new hardware to satisfy the ever increasing demand for cloud resources, and the main bottleneck in this process is power. Data centers consist of power devices with limited capacity and each demand for hardware (e.g., rack of servers) has a power requirement. The goal is to assign demands to power devices to fulfill their requirements while using the available power in the data centers efficiently. This allows cloud providers to maximize their return on investment on existing data centers before needing to incur large capital expenses for new data centers to accommodate additional demand. 







An important consideration that sets this demand assignment process apart from other applications is \emph{reliability}. Cloud users are promised a high availability of service which mandates that cloud capacity can only be unavailable for very short durations (between a few minutes and a few hours per year). As a result, assigning each demand to a single power device leads to an unacceptable level of risk; if that device fails, the capacity for the demand becomes unavailable, leading to potentially millions of dollars in costs for the provider and jeopardizing the cloud business model that is highly dependent on users' trust. To this end, power \emph{redundancy} is built into the assignment process.



\begin{figure}[h]
  \centering
  \scalebox{0.9}{
\begin{tikzpicture}
\draw (1,0) rectangle (2.2,3);
\draw[pattern=north west lines, pattern color=blue] (1,0) rectangle (2.2,1);
\draw[pattern=crosshatch dots, pattern color=blue] (1,1) rectangle (2.2,1.75);
\draw (3,0) rectangle (4.2,3);
\draw[pattern=north west lines, pattern color=blue] (3,0) rectangle (4.2,1);
\draw[pattern=dots, pattern color=blue] (3,1) rectangle (4.2,2.1);
\draw (5,0) rectangle (6.2,3);
\draw[pattern=crosshatch dots, pattern color=blue] (5,0) rectangle (6.2,0.75);
\draw[pattern=dots, pattern color=blue] (5,0.75) rectangle (6.2,1.85);
\node at (1.6, 0.5) {\contour{white}{\large $1$}};
\node at (1.6, 1.375) {\contour{white}{\large $2$}};
\node at (3.6, 0.5) {\contour{white}{\large $1$}};
\node at (3.6, 1.55) {\contour{white}{\large $3$}};
\node at (5.6, 0.375) {\contour{white}{\large $2$}};
\node at (5.6, 1.3) {\contour{white}{\large $3$}};
\node[black] at (1.6, -0.3) {\large $a$};
\node[black] at (3.6, -0.3) {\large $b$};
\node[black] at (5.6, -0.3) {\large $c$};
\node[black] at (3.6, -0.8) {\large Normal operations};
\draw (8.5,0) rectangle (9.7,3.5);
\draw[pattern=north west lines, pattern color=blue] (8.5,0) rectangle (9.7,1);
\draw[pattern=crosshatch dots, pattern color=blue] (8.5,1) rectangle (9.7,2.5);
\draw (10.5,0) rectangle (11.7,3.5);
\draw[pattern=north west lines, pattern color=blue] (10.5,0) rectangle (11.7,1);
\draw[pattern=dots, pattern color=blue] (10.5,1) rectangle (11.7,3.2);
\draw[dashed,color=gray] (12.5,0) rectangle (13.7,3);
\node[black] at (9.1, -0.3) {\large $a$};
\node[black] at (11.1, -0.3) {\large $b$};
\node[black] at (13.1, -0.3) {\large $c$};
\node[black] at (11.1, -0.8) {\large Failover due to device $c$};
\node at (9.1, 0.5) {\contour{white}{\large $1$}};
\node at (9.1, 1.75) {\contour{white}{\large $2$}};
\node at (11.1, 0.5) {\contour{white}{\large $1$}};
\node at (11.1, 2.1) {\contour{white}{\large $3$}};
\end{tikzpicture}
}
\caption{In normal operations (left), each demand (denoted with a different pattern) is assigned to two devices and gets half of its required power from each device. In the failover scenario where device $c$ has failed (right), the demands that were assigned to $c$ now get their full power from the remaining devices that may run at increased capacity.}
\label{fig:failover_example}
\end{figure}
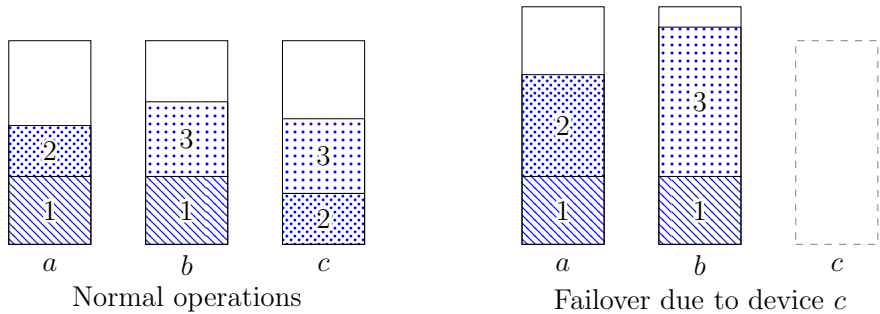

Specifically, each demand gets assigned to two power devices. In normal operations (no device failure), the demand obtains half of its required power from each device. If one of the devices fails, then the remaining device must provide the full power amount to the demand (see \Cref{fig:failover_example} for an example). In these failover scenarios, the remaining devices may run at an increased capacity temporarily to accommodate their increased load. The provider uses this time to take corrective actions, for instance, shut down certain workloads and reduce the power of others in order to bring the power utilization of each device back within its normal limits; see \cite{zhang2021flex} for more details on this process. Similar to \cite{zhang2021flex} we consider a single device failure at a time, since multiple devices failing simultaneously is highly unlikely. \knote{Is this last sentence ok or should we write it a different way?}



We introduce the \emph{Online Demand Scheduling with Failover} problem (\problem) to model this issue of assigning demands to power devices with redundancy. Formally, in this problem there are $m$ identical devices (or machines) and $n$ demands. Each  device has two capacities: a nominal capacity that is normalized to $1$ and a failover capacity $B \geq 1$. Each demand $j$ has some size $s_j \geq 0$, which for convenience is defined as its per-device power requirement 
(so the total power requirement of the demand is $2 s_j$).
The demands arrive online one-by-one and there is no knowledge about future demands. The goal is to irrevocably assign the arriving demands to pairs of devices (or edges, \replace{where we consider each device as a node}{}\knote{Ishai was saying that the edges were not well introduced, see if this helps}) satisfying:

\begin{enumerate}
	\item (Nominal Constraints) For every device $u$, its total load has to be at most 1, namely $L_u := \sum_{v \neq u} L_{uv} \leq 1$, where we define $L_{uv} = \sum_{j \rightarrow uv} s_j$ to be the total load on edge $uv$ (i.e., all demands assigned to the pair of devices $uv$).\label{nominal}
	\item(Failover Constraints) For every device $u$, we have $L_u + \max_{v \neq u} L_{uv} \leq B$\label{failover} \replace{(i.e., if a device $v\neq u$ fails, all demands assigned to $uv$ have to be supplied solely by $u$, which sees its load increased by the amount $L_{uv}$ that was formerly supplied to them by device $v$; the increased load has to fit the failover capacity $B$).}{}
\end{enumerate}

We assume that each demand size $s_j$ fits on a pair of devices by itself, so $s_j \in [0 , \min(1, B/2)]$. We are not allowed to reject demands, so the algorithm assigns arriving demands to the available devices until a demand cannot be scheduled, in which case the algorithm terminates. Our objective is to maximize the total size of all assigned demands (i.e., the utilization). 
We compare the algorithm against the optimal offline strategy that knows the demand sequence in advance (but still subject to the same no-rejection requirement). We use $\opt$ to denote the total utilization of this optimal offline strategy.

This problem has similarities with several classic packing problems. For example, in the Multiple Knapsack problem (and related problems such as Generalized Assignment~\cite{GAPShmoys}, AdWords~\cite{adWordsACM}, etc.) we are given a set of items each with a weight and size, and the goal is to select a subset of the items to pack in capacitated bins in order to maximize the total weight. However, one fundamental difference in our setting, besides the need to assign each demand to a pair of devices instead of a single device, is the failover constraint. Unlike in previously studied resource allocation problems, here the capacity constraints are not just determined by the total demand incident to a node, but rather they depend also on how the demands are arranged across its edges. See the next example.



\begin{example}
	 Consider an instance with 4 power devices $a$, $b$, $c$, $d$ with failover capacity $B = 1$, and where there are 6 demands of size $\frac{1}{4}$ that arrive sequentially. Suppose 4 demands have arrived so far and let us look at two potential assignment options: 
	
	
	
	\begin{itemize}
		\item \textbf{Bad assignment example.} Suppose we assign 2 demands to the pair $ab$ and 2 demands to the pair $cd$ (see \Cref{fig:example_failover_bad}). In this case, all devices still have available Nominal capacity, however the remaining two demands cannot be placed. To see this, assume we assign another demand to device $a$, say. The Nominal capacity for $a$ is satisfied. However, if device $b$ fails, then the total load on $a$ will become at least $\frac{5}{4}$ violating its Failover capacity. 
		\item \textbf{Good assignment example.} If instead we assign at most one demand to each device pair (see \Cref{fig:example_failover_good}), then all demands can be satisfied. In particular, if the first 4 demands are placed on pairs $ab$, $ac$, $bd$, $cd$ (solid edges in \Cref{fig:example_failover_good}), then the remaining two demands can be placed successfully on pairs $ad$ and $bc$ (dashed edges in \Cref{fig:example_failover_good}).
	\end{itemize}

	\tikzset{invisible/.style={minimum width=0mm,inner sep=0mm,outer sep=0mm}}
	\begin{figure}[h]
		\centering
		\begin{minipage}[b]{0.47\textwidth}
			\centering
			\noindent\begin{minipage}{\linewidth}
				\centering
				\begin{tikzpicture}[-,auto,scale = 1.2,
					thick,main node/.style={circle,draw,scale = 0.85},main node2/.style={circle,draw,fill=blue!60,scale=0.7},flow_a/.style ={blue!100}]
					\tikzstyle{edge} = [draw,thick,blue,-]
					\tikzstyle{cut} = [draw,very thick,-]
					\tikzstyle{flow} = [draw,line width = 1.5pt,->,blue!100]
					\node[main node] (1) at (0,1.2)  {a};
					\node[main node] (2) at (1.2,1.2) {b};
					\node[main node] (3) at (0,0) {c};	
					\node[main node] (4) at (1.2,0) {d};	
					\node[invisible, blue] at (0.6, 1.5) {$\frac{1}{4}+\frac{1}{4}$};
					\node[invisible, blue] at (0.6, -0.3) {$\frac{1}{4}+\frac{1}{4}$};
					\path[edge]	(1) edge (2);
					\path[edge]	(3) edge (4);
				\end{tikzpicture}
			\end{minipage}
			\caption{Example of bad assignment.}
			\label{fig:example_failover_bad}
		\end{minipage}
		\hfill
		\begin{minipage}[b]{0.47\textwidth}
			\centering
			\noindent\begin{minipage}{\linewidth}
				\centering
				\begin{tikzpicture}[-,auto,scale = 1.2,
					thick,main node/.style={circle,draw,scale = 0.85},main node2/.style={circle,draw,fill=blue!60,scale=0.7},flow_a/.style ={blue!100}]
					\tikzstyle{edge} = [draw,thick,blue,-]
					\tikzstyle{cut} = [draw,very thick,-]
					\tikzstyle{flow} = [draw,line width = 1.5pt,->,blue!100]
					\node[main node] (1) at (0,1.2)  {a};
					\node[main node] (2) at (1.2,1.2) {b};
					\node[main node] (3) at (0,0) {c};	
					\node[main node] (4) at (1.2,0) {d};	
					\node[invisible, blue] at (0.6, 1.5) {$\frac{1}{4}$};
					\node[invisible, blue] at (0.6, -0.3) {$\frac{1}{4}$};
					\node[invisible, blue] at (-0.2, 0.6) {$\frac{1}{4}$};
					\node[invisible, blue] at (1.4, 0.6) {$\frac{1}{4}$};
					\node[invisible, blue] at (0.3, 0.6) {$\frac{1}{4}$};
					\node[invisible, blue] at (0.9, 0.6) {$\frac{1}{4}$};
					\path[edge]	(1) edge (2);
					\path[edge]	(1) edge (3);
					\path[edge]	(1) edge[dashed] (4);
					\path[edge]	(2) edge[dashed] (3);
					\path[edge]	(2) edge (4);
					\path[edge]	(3) edge (4);
				\end{tikzpicture}
			\end{minipage}
			\caption{Example of good assignment.}
			\label{fig:example_failover_good}
		\end{minipage}
	\end{figure}
	
\end{example}

The above example suggests that due to the Failover constraints we should ``spread out'' the demands by not putting too many demands on one edge, because if one of its endpoints fails then this edge can have a large contribution to the Failover constraint of the other endpoint. However, there is a danger in spreading out the demands too much and not leaving enough devices free. 

\begin{example}
	Consider again the same 4 power devices $a$, $b$, $c$, $d$ with failover capacity $B = 1$. Now, there are 7 demands; the first 6 have a small size $\e > 0$ and the last demand has size 0.5. Assume the first 6 demands have arrived and let us look at two potential assignment options.
	
	
	
	\begin{itemize}
		\item \textbf{Bad assignment example.} Suppose we assign one demand of size $\epsilon$ per device pair (see \Cref{fig:example_failover_bad_eps}). In this case, the remaining demand of size 0.5 cannot be placed, as the Failover capacities would be exceeded.
		\item \textbf{Good assignment example.} If instead we group the first 6 demands on a single edge (see \Cref{fig:example_failover_good_eps}), then all demands can be fulfilled by assigning the last demand on a disjoint edge (dashed edge of \Cref{fig:example_failover_good_eps}).
	\end{itemize}

	\tikzset{invisible/.style={minimum width=0mm,inner sep=0mm,outer sep=0mm}}
	\begin{figure}[h]
		\centering
		\begin{minipage}[b]{0.47\textwidth}
			\centering
			\noindent\begin{minipage}{\linewidth}
				\centering
				\begin{tikzpicture}[-,auto,scale = 1.2,
					thick,main node/.style={circle,draw,scale = 0.85},main node2/.style={circle,draw,fill=blue!60,scale=0.7},flow_a/.style ={blue!100}]
					\tikzstyle{edge} = [draw,thick,blue,-]
					\tikzstyle{cut} = [draw,very thick,-]
					\tikzstyle{flow} = [draw,line width = 1.5pt,->,blue!100]
					\node[main node] (1) at (0,1.2)  {a};
					\node[main node] (2) at (1.2,1.2) {b};
					\node[main node] (3) at (0,0) {c};	
					\node[main node] (4) at (1.2,0) {d};	
					\node[invisible, blue] at (0.6, 1.5) {$\epsilon$};
					\node[invisible, blue] at (0.6, -0.3) {$\epsilon$};
					\node[invisible, blue] at (-0.2, 0.6) {$\epsilon$};
					\node[invisible, blue] at (1.4, 0.6) {$\epsilon$};
					\node[invisible, blue] at (0.3, 0.6) {$\epsilon$};
					\node[invisible, blue] at (0.9, 0.6) {$\epsilon$};
					\path[edge]	(1) edge (2);
					\path[edge]	(1) edge (3);
					\path[edge]	(1) edge (4);
					\path[edge]	(2) edge (3);
					\path[edge]	(2) edge (4);
					\path[edge]	(3) edge (4);
				\end{tikzpicture}
			\end{minipage}
			\caption{Example of bad assignment.}
			\label{fig:example_failover_bad_eps}
		\end{minipage}
		\hfill
		\begin{minipage}[b]{0.47\textwidth}
			\centering
			\noindent\begin{minipage}{\linewidth}
				\centering
				\begin{tikzpicture}[-,auto,scale = 1.2,
					thick,main node/.style={circle,draw,scale = 0.85},main node2/.style={circle,draw,fill=blue!60,scale=0.7},flow_a/.style ={blue!100}]
					\tikzstyle{edge} = [draw,thick,blue,-]
					\tikzstyle{cut} = [draw,very thick,-]
					\tikzstyle{flow} = [draw,line width = 1.5pt,->,blue!100]
					\node[main node] (1) at (0,1.2)  {a};
					\node[main node] (2) at (1.2,1.2) {b};
					\node[main node] (3) at (0,0) {c};	
					\node[main node] (4) at (1.2,0) {d};	
					\node[invisible, blue] at (0.6, 1.5) {$6\epsilon$};
					\node[invisible, blue] at (0.6, -0.3) {$0.5$};
					\path[edge]	(1) edge (2);
					\path[edge]	(3) edge[dashed] (4);
				\end{tikzpicture}
			\end{minipage}
			\caption{Example of good assignment.}
			\label{fig:example_failover_good_eps}
		\end{minipage}
	\end{figure}
	
\end{example}

Taking these two examples together, we see that there is a delicate balance between spreading demands out across edges to minimize their impact in failover scenarios and leaving enough devices open for future demands, as to not prematurely end up with an unassignable demand. 

\subsection{Our results} 

	\mnote{Take another pass on this subsec}
	We start by considering the \problem problem in the worst-case and design a deterministic algorithm with competitive ratio $\approx \frac{1}{2}$. Since no deterministic algorithm can be better than $\frac{1}{2}$-competitive (\Cref{thm:upperBoundDet} in \Cref{sec:onWCLB}), this result is almost best possible.\mnote{One possibility is to mention in passing the randomized bound (and add to app)} \replace{(For the special case where demand sizes are small, we adapt our algorithm to obtain an improved competitive ratio, see \Cref{thm:small} in \Cref{sec:onWCSsmall}.)}{}
	
\begin{restatable}{thm}{thmMainWorst}\label{thm:main}
		There is a deterministic \replace{poly-time}{} online algorithm for \problem in the worst-case model that has competitive ratio at least $\frac{1}{2} - O(\frac{1}{m^{1/3}})$,\footnote{Throughout the paper we use $O(x)$ to mean ``$\le cst \cdot x$'' for some constant $cst$ independent of $x$.} where $m$ is the number of devices.
\end{restatable}






\mnote{I think I prefer the older store without the ``small sizes'' result here: we were basically saying it is impossible to do better then $\frac{1}{2}$, which comes across stronger.}
	A $\frac{1}{2}$-competitive solution may, roughly speaking, underutilize by a factor of $\frac{1}{2}$ the available power; in the context of big cloud providers, this inefficiency translates to substantial capital expenses due to the extra data centers required to accommodate the demands.
	\mnote{Konstantina, I need your help with this type of text :)}\knote{changed this a little} Since such losses are unavoidable in the worst-case model\mnote{But $\frac{1}{2}$ is only for deterministic algorithms\ldots}, we consider the \problem problem in the \emph{stochastic arrival model}. Here the demand sizes are drawn i.i.d. from an unknown distribution $\mu$ supported on $[0, \min(1, B/2)]$. 
	
	We show that in this stochastic model it is possible to obtain \emph{sublinear additive regret}. This means that as $\opt$ (or, equivalently, the number of devices) grows, the multiplicative competitive ratio of our algorithm goes to $1$. \replace{}{Note that this regime is of particular interest for large cloud providers.}

\begin{restatable}{thm}{thmMainStoch} \label{thm_main_stoch}
	For the \problem problem in the stochastic arrival model, there is a poly-time algorithm that achieves utilization at least $\OPT - O(\OPT^{5/6} \log \OPT)$ with probability $1 - O(\frac{1}{m})$. 
\end{restatable}

As a subroutine of this algorithm, we need to solve the natural \emph{offline minimization} variant of demand scheduling with failover: Given a collection of demands, minimize the number of devices needed to assign all demands satisfying the Nominal and Failover constraints. We also design an (offline) algorithm with sublinear additive regret for this problem (\Cref{sec_offline_min_machines}). 


\subsection{Technical Overview}

We illustrate the main technical challenges in the \problem problem in both the worst-case and stochastic models, as well as in the offline minimization subproblem needed for the latter.

\paragraph{Online Worst-Case {\normalfont (\Cref{sec:onWCSNR})}.} 
	The examples from \Cref{fig:example_failover_good} and \ref{fig:example_failover_good_eps} show that the main difficulty is dealing with the trade-off between spreading out the demands, which allows for a better use of the failover budgets, and co-locating demands on fewer edges, keeping some edges free for future big demands. 
	 
	To effectively strike this balance and get near optimal guarantees, the main idea is to group demands based on their sizes using intervals $I_k$ and schedule each group separately on cliques of size $k$. That is, we will ``open'' a set of $k$ unused devices and assign the demands in $I_k$ only to the edges between these devices (opening new $k$-cliques as needed). Interestingly, we  assign at most one demand per edge of the clique (other than for tiny demands, which are handled separately). This means the algorithm tries to co-locate demands in controlled \emph{regions}
	, which allows for the right use of the failover budgets. \mnote{Is it making sense? Similar to the prev version, just a bit longer}
\mnote{Maybe mention that it is easy to get a $O(1)$-approx: assign anything bigger than $1/4$ (or something) to its own edge, the other jobs in any machine with occupation less than $1/4$. So our contribution is to get the tight (for deterministic algorithms) (2+o(1))-approx}

\paragraph{Online Stochastic Arrivals {\normalfont (\Cref{sec:stoch})}.}  
\mnote{This is tough to give intuition, see if the whole thing makes sense}First, note that because demands are i.i.d. from a distribution with bounded support, the total utilization of the first $\ell$ demands grows as \replace{$\ell \cdot \E_{S \sim \mu} S$}{$\mathbb{E}[\mu] \cdot \ell$}. Thus, it suffices to show that our algorithm ``survives'' for as many demand arrivals as possible without needing to reject one due to lack of space. Our approach is to try and assign prefixes of arrivals to the (approximately) minimum number of devices possible. This ensures that if our algorithm fails due to needing more than $m$ devices to feasibly assign another demand, then $\opt$ will fail shorty after.

Our algorithm is based on a learn-and-pack framework, where we use knowledge of the first $\ell$ arrivals to compute a good template assignment for the next $\ell$ arrivals. To compute this template, we need a subroutine that (approximately) solves the offline minimization subproblem mentioned above. Concretely, we run the subroutine on the realized sizes of the first $\ell$ arrivals, which gives a possible assignment of these demands into, say $m'$ unused devices. We use the ``slots'' of this possible assignment as a template to assign the future $\ell$ demands by employing the \emph{online monotone matching} process of Rhee-Talgrand~\cite{rhee1993line}: For each future arrival, we assign it to a (carefully-chosen) open slot in the template that has a larger size -- if we cannot find such an open slot, then we assign this demand to its own disjoint edge (using $2$ more devices).

It is known that this matching process leaves $o(\ell)$ unmatched demands with high probability. Further, our offline minimization subroutine has sublinear additive regret, that is, it uses only $o(\ell)$ more devices that the optimal offline assignment. \replace{Since these losses are sublinear in the prefix size, it seems that by repeating this process together with doubling the prefix size we should obtain a final sublinear regret guarantee.}{}

\replace{But there is still a major issue: This strategy uses \emph{disjoint} sets of devices to fulfill the first $\ell$ demands and the next $\ell$ demands (for each doubling $\ell$). But this is possibly very wasteful: even using the optimal assignment for each of these $\ell$ demands separately may require many more devices (up to double) compared to reusing the leftover space from the first batch of $\ell$ demands for the next batch (i.e. assigning the batches to a common set of devices). Wasting a constant fraction of devices would lead to the unwanted constant-competitive loss. To overcome this, we show that $M_\ell$, the minimum number of devices to assign $\ell$ i.i.d. demands, is approximately linear in $\ell$ (\Cref{sec:asConverge}), e.g. $M_\ell + M_\ell$ (assigning batches separately) is approximately $M_{2\ell}$ (assigning them together). This is a non-trivial task (another Rhee-Talagrand paper~\cite{rhee1989StochasticConv} is entirely devoted to doing this for the simpler Bin Packing problem). Perhaps surprisingly, our proof relies on our algorithm for the offline device minimization problem, which is LP-based. The crucial property is that the optimal LP value doubles if we duplicate the items on its input, which (with additional probabilistic arguments) translates into the additivity of $M_\ell$.}{}



\paragraph{Offline Minimization {\normalfont (\Cref{sec_offline_min_machines})}.}
Our algorithm for offline minimization of the number of devices needed to fulfill a set of demands is based on a configuration LP inspired by the classic Gomory-Gilmore LP\knote{Cite or not needed?} for the Bin Packing problem. Consider a fixed assignment of demands to some number of devices. \replace{We}{Ideally, we}\knote{The "ideally" may not be reading that well, see how to change} want to interpret each device as a configuration, which captures the arrangement\knote{should this be assignment instead of arrangement? Because we don't say exactly how they are arranged in the config} of demands on this device's edges. Our LP will minimize the number of configurations needed in order to assign all demands.

There is a tension between two issues in this approach. First, the Failover constraint depends not only on the subset of demands on this device's edges, but also how they are arranged within these edges (because the most-loaded edge contributes to the Failover constraint). This suggests that a configuration should not only specify a subset of demands, but also have enough information about the edge assignment to control the most-loaded edge. Second, each demand must be assigned to a pair of devices rather than a single device, so our configurations are not ``independent'' of each other. Thus, we need to ``match'' configurations to ensure that a collection of configurations can be realized in an edge assignment. In summary, our configurations should be expressive enough to capture the Failover constraints, but also simple enough so that we can actually realize them in an actual assignment.

Our solution to this is to define a configuration to be a subcollection, say $C$, of demands satisfying $\sum_{s \in C} s \leq 1$ (the Nominal constraint) and $\sum_{s \in C} s + \max_{s \in C} s \leq B$ (a relaxed Failover constraint). Note that this notion of configuration does not capture the arrangement of the demands $C$ across a device's edges -- we assume the best case  that every demand is on its own edge to minimize their impact in failover scenarios. It is not clear that there even exists a near-optimal assignment that assigns at most one demand per edge, let alone that we can obtain one from the LP solution. However, our LP post-processing procedure will show that -- by opening slightly more devices -- we can match configurations of this form to realize them in a near-optimal assignment.

\subsection{Related work}

\mnote{Try to shorten/make it crisper this sec} Despite a vast literature on assignment-type problems, none of the ones considered addresses the main \replace{issue of redundancy}{}, modeled in the \problem problem. Arguably the Coupled Placement~\cite{kSidedPlacement} problem is the closest to \problem. Given a bipartite graph with capacities at the nodes and a set of jobs, the goal is to assign a subset of the jobs to the edges of the graph to maximize the total value (each assigned job gives a value that also depends on its assigned edge), while respecting the capacity of the nodes (each assigned job consumes capacity from its edge's endpoints).
~\cite{kSidedPlacement} gives a $\frac{1}{15}$-approximation to the offline version of this problem (see also~\cite{demandMatching}). While this problem involves the allocation of jobs to a pair of nodes (albeit on a bipartite graph) and has the additional difficulty that the value and consumption of a job depends on which pair of nodes it is assigned, it does not have any Failover type constraints, a crucial component of our problem. 


As already mentioned, several classic assignment problems are related to ours, such as the Multiple Knapsack~\cite{multipleKnapsackChekuri}, Generalized Assignment (GAP)~\cite{GAPShmoys}, and AdWords problem~\cite{adWordsACM}. The latter is the closest to our problem: there are $m$ bins (i.e. advertisers) of different capacities, and jobs (i.e. keyword searches) that come one-by-one and need to be assigned to the bins; each assignment consumes some of the bin's capacity and incurs an equal amount of value (i.e. bid). The goal is to find an online assignment that maximizes the total value obtained subject to the bin capacity constraints. In the worst-case model,~\cite{adWordsACM} provides an algorithm with the optimal competitive ratio of $1-\frac{1}{e}$ (under the assumption that the bids are arbitrarily smaller than the capacities). In the stochastic model, if job rejections are allowed~\cite{guptaMol} obtains competitive ratio $\sqrt{(\log m)/B}$, where $B$ is the smallest capacity/job-size ratio. Despite the similarities, this problem does not consider critical aspects of our problem, namely the need to assign a job/demand to a pair of bins/devices and the Failover constraints. 

There is also a large literature on survivable network design problems, where failures in the network are explicitly considered~\cite{survivableSurvey}, but the nature of the problems is quite different from our assignment problem \replace{as the focus there is typically on routing flows}{}. 

Finally, a problem related to our device minimization problem\mnote{Improve name, give acronym or something}, and from which we borrow some tools and techniques, is Bin Packing. Here jobs of different sizes need to be assigned to a minimum number of bins of size 1. In the offline setting the best result is an additive $+O(\log \opt)$ approximation due to Hoberg and Rothvoss~\cite{logBinPacking}. In the online setting where jobs come one-by-one, in the worst-case model the current best competitive ratio is $\approx 1.57829$~\cite{binPackingWorstCaseESA}. In the stochastic model,~\cite{rhee1993line} obtains an additive $+ O(\sqrt{\opt} \cdot \log^{3/4} \opt)$ sublinear approximation; see also~\cite{csirik2006sum,binPackingVarun,binPackingKnownT} for improvements under different assumptions.


	

\ifx \hasmain \undefined

\usepackage{fullpage}
\usepackage[pdftitle={Tile},
            pdfauthor={Author},
            pdfkeywords={Keywords}]{hyperref}

\input{preamble}

\begin{document}

\fi
	
\section{\problem Problem in the Online Worst-Case Model} \label{sec:onWCSNR}

In this section we consider the \problem in the online worst-case model. We design an algorithm that achieves competitive ratio $\approx \frac{1}{2}$ in this setting (restated from the introduction).


\thmMainWorst*

Recall that in \Cref{app:onWCS} we also show the almost matching upper bound of $\frac{1}{2}$ on such competitive ratio, and design another  algorithm whose competitive ratio approaches 1 as the size of the largest demand goes to 0. To convey the main ideas more clearly, here we focus only on \Cref{thm:main}.

	\subsection{Algorithm} 
	
	\knote{Copy-paste from technical - rewrite shorter}
	\replace{As suggested in the technical overview, our algorithm will group demands by size, and assign each group of demands to sub-cliques of an appropriate size.}{}

	To make this precise, set in hindsight $L := m^{1/3}$ and for $k = 2, \ldots, L-1$ define the interval $$I_k := \bigg(\min\bigg\{\frac{1}{k}, \frac{B}{k+1}\bigg\}~,~ \min\bigg\{\frac{1}{k-1}, \frac{B}{k} \bigg\}  \bigg].$$ (Notice there is no $k=1$, because the upper limit of $I_2$ is the max size of a demand.) This definition ensures that it is feasible to assign one demand of such size to each edge of a $k$-clique, as we argue in the next subsection. Also define the interval of small sizes $$I_{\ge L} := \bigg[0, \min\bigg\{\frac{1}{L-1}, \frac{B}{L}\bigg\} \bigg].$$
	
	The algorithm is then the following:

\begin{algorithm}[H]
\small
\caption*{\textbf{FailoverWostCase:}}
\begin{algorithmic}[1]
	\itemsep=1.5ex
	
	\vspace{1ex}
		\State When a demand arrives, determine the interval $I_k$ (or $I_{\ge L}$) that it belongs to based on its size. 
		
		\State If it belongs to an interval $I_k$ with $k \in \{2,\ldots,L-1\}$, assign the demand to any ``empty'' edge (i.e. that has not received any demands) of a $k$-clique opened for $I_k$. If no such edge exists, then open a new $k$-clique for $I_k$. 

		\State Otherwise it belongs to $I_{\ge L}$, so assign it to an edge of one of its $L$-cliques using \emph{first-fit} (so here we \textbf{can assign multiple demands to the same edge}) making sure that the total load on each edge is at most $\min\{\frac{1}{L-1}, \frac{B}{L}\}$. By first-fit we mean that the edges of the $I_{\ge L}$ cliques are arbitrarily ordered and the demand is assigned to the first possible edge. Open a new $L$-clique for $I_{\ge L}$ if need be.
			
		\State If the demand cannot fit in the appropriate clique and it is not possible to open a new clique (i.e. there are not enough unused machines to form a clique of the desired size), then stop.
\end{algorithmic}
\end{algorithm}


	\subsection{Analysis} 
	
	\mnote{Maybe move the whole analysis to the appendix?}
	We first quickly verify that the assignment done by the algorithm is feasible, i.e. satisfies the Nominal and Failover constraints. Consider a node/machine $u$ on an $I_k$ clique opened by the algorithm (for machines in an $I_{\ge L}$ clique the argument is the analogous). For the Nominal capacity constraint: Every demand assigned to $u$ is actually assigned to one of the $k-1$ edges in this clique incident on $u$; each such edge receives at most 1 demand from $I_k$ (and no other demands), so using the upper limit of this interval we see that $u$ receives total size at most $(k-1)\cdot\min\{\frac{1}{k-1},\frac{B}{k}\} \le 1$, so within its Nominal capacity. For the Failover capacity: in a failover scenario one of these $(k-1)$ demands has ``both ends'' assigned to $u$, so the total size it receives is now $k \cdot \min\{\frac{1}{k-1},\frac{B}{k}\} \le B$, so within the Failover capacity. Hence the algorithm produces a feasible assignment.
	
	\medskip 
	Now we show that the value obtained by the algorithm is at least $\big(\frac{1}{2} - O(\frac{1}{m^{1/3}})\big)\opt$. The idea is to show that for (essentially) each clique opened by the algorithm, we get on average value at least $\approx \frac{1}{4}$ per vertex. Given that each node has Nominal capacity 1 and each demand must be scheduled on two nodes, $\opt$ can only get at most $\frac{1}{2}$ value from each node on average, so this shows that our algorithm is a $\approx \frac{1}{2}$-approximation. 
	However, there are two exceptions where we may get less than $\approx \frac{1}{4}$ per vertex on average. The first is the last clique for each $I_k$, which may not be ``fully used'' (but by setting $L$ appropriately there are not too many nodes involved in this loss). More importantly, \replace{the second exception is}{are} the ``big items'' $I_2$, which may not allow us to get average value $\frac{1}{4}$ per node (e.g. when the failover is $B=1$, a demand of size $\frac{1}{3} + \e$ falls in the group $I_2$ and is put by itself on an edge, giving value $\frac{1}{6} + \frac{\e}{2} \ll \frac{1}{4}$ per node used). However, in this case we show that we can obtain a stronger upper bound for these demands for \opt.
	

	We now make this precise. Assume throughout that the algorithm has stopped before the end of the input (else it scheduled everything, so it is $\opt$). We account for the value obtained on each type of clique separately.

	\paragraph{Cliques for $I_{\ge L}$.} We will use two observations: 
	\begin{itemize}
	    \item When the algorithm opens a new $I_{\ge L}$ clique, every edge of the previous $I_{\ge L}$ cliques has some demand assigned to it.
	    \item Across all $I_{\ge L}$ cliques, out of all edges with some demand assigned to them, at most one can have total size assigned to it less than  $\alpha := \frac{1}{2} \min\big\{\frac{1}{L-1}, \frac{B}{L} \big\}$ (i.e. half of its ``capacity'').
	\end{itemize}
	
	Both observations stem from the first-fit strategy to assign these demands. In particular, the algorithm will only open a new clique when a demand in $I_{\ge L}$ does not fit in the edges of the existing cliques, implying that all of these edges already have some demand assigned; this shows the first statement.	For the second statement, by contradiction assume that at some point there are at least two edges on $I_{\ge L}$ cliques with total load less than  $\alpha$. Then the first demand that was assigned to the last such edge has size less than $\alpha$. But this means that it could have been assigned to an earlier edge with load less than $\alpha$, contradicting the first-fit procedure. \mnote{We can possibly make these observations into a lemma and send this paragraph/proof to the appendix to save space (but not much)}
		
	Let  $c_{\ge L}$ be the total number of $I_{\ge L}$ cliques that the algorithm opened, and $m_{\ge L} := c_{\ge L} \cdot L$ the number of nodes/machines associated with those cliques. Combining the above two observations, at the end of the execution either: (i) every edge of the first $c_{\ge L}-1$ of these cliques has load at least $\alpha$ or; (ii) all but one edge in the first $c_{\ge L} - 1$ cliques has load at least $\alpha$ and some edge of the last $c_{\ge L}$-th (e.g., the one that ``opened'' it) has load at least $\alpha$. In both cases, the total size of demands assigned by the algorithm to the edges of these cliques is at least
	\begin{align}
	    (c_{\ge L} - 1) \cdot {L \choose 2} \cdot \alpha \,&=\, (c_{\ge L} - 1) \cdot \frac{L}{4}\cdot \min\bigg\{1, \frac{(L-1) B}{L}\bigg\} \nonumber\\ & \geq  (c_{\ge L} - 1) \cdot \frac{L}{4} \bigg( 1 - \frac{1}{L}\bigg) \,=\, m_{\ge L} \cdot \frac{1}{4} \bigg(1 - \frac{1}{L}\bigg) - O(L), \label{eq:L}
	\end{align}
yielding roughly average value $\frac{1}{4}$ from each node of these cliques, as claimed.

	\paragraph{Cliques for $I_{k}$, for $k \ge 3$.} Consider any clique for $I_k$ except the last one to be opened. All edges of this clique have some demand from $I_k$ assigned to it; given the lower limit for this interval, this means that the algorithm has assigned to each such clique total size\mnote{Sometimes using total size/demand/value/utilization, see if ok for the reader} at least 
	\begin{align}
		{k \choose 2}\cdot  \min\bigg\{\frac{1}{k}, \frac{B}{k+1}\bigg\} \,=\, \frac{k}{2} \cdot \min\bigg\{\frac{k-1}{k}, \frac{B (k-1)}{k+1}\bigg\}. \nonumber 
	\end{align}
	Since $k \ge 3$ and $B \ge 1$, the right-hand side is at least $\frac{k}{4}$. Letting again $c_k$ denote the number of cliques for $I_k$ that the algorithm opens and $m_k$ the corresponding number of nodes/machines, we can count the total value of all but the last $I_k$ clique and we see that the algorithm has assigned to them total size at least 
	\begin{align}
	(c_k -1) \cdot \frac{k}{4} \,=\, m_k \cdot \frac{1}{4} - O(k). \label{eq:k1}
	\end{align}
	
	\paragraph{Cliques for $I_2$.} (Recall that there is no $k=1$, so this is the last case to consider.) Given the lower limit of the interval $I_2$, each $I_2$ clique (which being a $2$-clique is just an edge) has a demand of size at least $\min\{\frac{1}{2}, \frac{B}{3}\}$ assigned to it. So the algorithm assigns total size at least $m_2 \cdot \min\{\frac{1}{4}, \frac{B}{6}\}$ to these $I_2$ cliques, where $m_2$ is the number of nodes in these cliques. 
	
	\paragraph{Total value of Alg.} Since we assumed that the algorithm stops at some point, it means that it could not open more cliques. This means that all but at most $L-1$ nodes belong to one such clique (the worst case is that it tried to open an $L$-clique but could not), so $m_{\ge L} + \sum_{k=3}^{L-1} m_k + m_2 \ge m - L$. Then adding the above estimates for the values obtained on each type of clique, we see that the algorithm gets total value at least
	\begin{align*}
		\alg ~&\ge~ \frac{1}{4} \bigg(1 - \frac{1}{L}\bigg) \cdot \bigg(m - m_2 - L\bigg) - O(L^2) + m_2 \cdot \min\bigg\{\frac{1}{4}, \frac{B}{6}\bigg\}\notag\\
		&= \frac{1}{4} \cdot \big(m - m_2\big) + m_2 \cdot \min\bigg\{\frac{1}{4}, \frac{B}{6}\bigg\}- O(m^{2/3}) 
	\end{align*}
	where the last inequality uses the fact that $L = m^{1/3}$. 
	

	Notice that if the minimum in the last line is $\frac{1}{4}$, then we obtain $\alg \ge \big(\frac{1}{2} - O(\frac{1}{m^{1/3}})\big) \opt$ as desired  (recall $\opt \le \frac{m}{2}$ since each machine has Nominal capacity 1 and each demand is assigned to two machines). \textbf{So assume this is not the case, namely $B < \frac{3}{2}$.} Under this assumption 
	\begin{align}
		\alg ~&\stackrel{with~ass.}{\ge}~ \frac{1}{4} \cdot \big(m - m_2\big) + \frac{B}{6} \cdot m_2 - O(m^{2/3}) \label{eq:finalAlgAss}
	\end{align}
	
	\paragraph{Value of \OPT.} We analyze \OPT again under the assumption $B < \frac{3}{2}$. The Failover constraints also ensure that in order to accommodate the demand from $I_2$ in case of failure, any node that receives a demand from $I_2$ can have total size assigned to it a most
	%
	\begin{align*}
		B - \min\bigg\{\frac{1}{2}, \frac{B}{3} \bigg\} \stackrel{with~ass.}{=} \frac{2B}{3},
	\end{align*}
	the last equation due to the assumption $B < \frac{3}{2}$. For all other nodes, $\OPT$ can assign at most size 1 per node due to the Nominal capacity constraint. Let $m_2^{\OPT}$ be the number of nodes where $\OPT$ schedules a demand from $I_2$. Again, since the size of each demand is counted towards the Nominal capacity of two nodes, the total size scheduled by $\OPT$ is 
	\begin{align}
	\OPT ~&\le~ \frac{1}{2} \left(m_2^{\OPT} \cdot \frac{2B}{3} +  (m- m_2^{\OPT}) \cdot 1  \right) \,=\, \frac{1}{2} \cdot (m- m_2^{\OPT}) + \frac{B}{3} \cdot m_2^{\OPT}  
	\label{eq:OPTAss}
	\end{align}

	
	%
	
	Notice that since every demand in $I_2$ has size $> \min\{\frac{1}{2}, \frac{B}{3}\} \ge \frac{1}{3}$, the Failover constraints ensure that in \OPT (as well as in our algorithm) the demands from $I_2$ that are scheduled form a matching, i.e. no 2 such demands can share a node/machine. So $m^{\OPT}_2$ (resp. $m_2$) is just twice the number of $I_2$ demands scheduled by $\OPT$ (resp. our algorithm). Moreover, both \alg and \OPT schedule a prefix of the instance. Since \OPT gets at least as much value as \alg, it means that it scheduled a prefix that is at least as long; in particular it schedules at least as many $I_2$ demands as our algorithm. Together these observations imply that that $m_2^{\OPT} \ge m_2$.	Then given inequalities \eqref{eq:finalAlgAss} and \eqref{eq:OPTAss}, under the assumption $B < \frac{3}{2}$ we obtain that $\alg \ge \big(\frac{1}{2} - O(\frac{1}{m^{1/3}})\big)\opt$ as desired. This concludes the proof of Theorem \ref{thm:main}.

\ifx \hasmain \undefined
	\end{document}
\fi

	\section{Sublinear Additive Regret in the Stochastic Model} \label{sec:stoch}
	
	We now consider \problem in the online stochastic model, where, instead of being adversarial, the size $S_t$ of each demand now comes independently from an unknown distribution $\mu$ over $[0,\min\{1,\frac{B}{2}\}]$. Again, at time $t$ the algorithm observes the size $S_t$ of the current demand and irrevocably assigns it to two of the $m$ machines. We still use $\OPT =  \OPT(S_1,\ldots,S_n)$ to denote the value \replace{of}{} (sum of the sizes scheduled by) the optimal strategy, which is now a random quantity. 
	
	Our  main result is algorithm \failoverstochastic that achieves a sublinear additive loss compared to $\opt$ in this stochastic model (restated from the introduction for convenience). 
	
	

\thmMainStoch*

	
	The algorithm relies on a learn-and-pack approach that uses previously seen items to compute a \emph{template} for packing the next items. This process is performed in rounds. Each round starts by assigning the first demand of the round on a pair of (empty) machines. Then, we iteratively create a template for the first $n_k := 2^k$ items of the round, which we use to schedule the next $n_k$ items. When the number of machines needed for the template (along with some slack) exceeds the number of available machines, the current round terminates and the next round begins. The next round maintains no knowledge of the previous demands; it only takes as input the number of empty machines $\tilde{m}$ which it is allowed to use. 
	A schematic overview of this process is presented in Figure \ref{fig:stoch_alg_illustration}. 

	\tikzset{invisible/.style={minimum width=0mm,inner sep=0mm,outer sep=0mm}}
	\begin{figure}[h]
		\centering
			\noindent\begin{minipage}{\linewidth}
				\centering
				\begin{tikzpicture}[-,auto,scale = 1.2,
					thick,main node/.style={circle,fill=blue,scale = 0.2},main node2/.style={circle,draw,fill=blue!60,scale=0.7},flow_a/.style ={blue!100}]
					\tikzstyle{edge} = [draw,thick,blue,-]
					\tikzstyle{cut} = [draw,very thick,-]
					\tikzstyle{flow} = [draw,line width = 1.5pt,->,blue!100]
					\node[main node,label=above:{$S_1$}] (1) at (0,0)  {$S_1$};
					\node[main node,label=above:{$S_2$}] (2) at (0.8,0) {$S_2$};
					\node[main node,label=above:{$S_3$}] (3) at (1.6,0) {$S_3$};	
					\node[main node,label=above:{$S_4$}] (4) at (2.4,0) {$S_4$};	
					\node[main node,label=above:{$S_5$}] (4) at (3.2,0) {$S_4$};	
					\node[main node,label=above:{$S_6$}] (4) at (4.0,0) {$S_4$};	
					\node[main node,label=above:{$S_7$}] (4) at (4.8,0) {$S_4$};	
					\node[main node,label=above:{$S_8$}] (4) at (5.6,0) {$S_4$};	
					\node[main node,label=above:{$S_9$}] (4) at (6.4,0) {$S_4$};	
					\node[main node,label=above:{$S_{10}$}] (4) at (7.2,0) {$S_4$};	
					\node[main node,label=above:{$S_{11}$}] (4) at (8.0,0) {$S_4$};	
				    \node[main node,label=above:{$S_{12}$}] (4) at (8.8,0) {$S_4$};	
					\node[main node,label=above:{$S_{13}$}] (4) at (9.6,0) {$S_4$};	
					\node[main node,label=above:{$S_{14}$}] (4) at (10.4,0) {$S_4$};	
					\node[main node,label=above:{$S_{15}$}] (4) at (11.2,0) {$S_4$};	
					\node[main node,label=above:{$S_{16}$}] (4) at (12.0,0) {$S_4$};	
					\node[main node,label=above:{$S_{17}$}] (4) at (12.8,0) {$S_4$};				
                    \tikzset{
                        ncbar angle/.initial=-90,
                        ncbar/.style={
                            to path=(\tikztostart)
                            -- ($(\tikztostart)!#1!\pgfkeysvalueof{/tikz/ncbar angle}:(\tikztotarget)$)
                            -- ($(\tikztotarget)!($(\tikztostart)!#1!\pgfkeysvalueof{/tikz/ncbar angle}:(\tikztotarget)$)!\pgfkeysvalueof{/tikz/ncbar angle}:(\tikztostart)$)
                            -- (\tikztotarget)
                        },
                        ncbar/.default=0.5cm,
                    }
				\tikzset{square bottom brace/.style={ncbar=0.1cm}}
	            \draw [thick] (-0.3,-0.2) to [square bottom brace ] (0.3,-0.2);
     		    \node [fill=white,inner sep=1pt] at (0, -0.3) {$n_0$};
                \tikzset{square bottom brace/.style={ncbar=0.1cm}}
	            \draw [thick] (-0.3,-0.45) to [square bottom brace ] (1.1,-0.45);
			    \node [fill=white,inner sep=1pt] at (0.4, -0.55) {$n_1$};
			     \draw [thick] (-0.3,-0.7) to [square bottom brace ] (2.7,-0.7);
			    \node [fill=white,inner sep=1pt] at (1.2, -0.8) {$n_2$};
			    \draw [thick] (-0.3,-0.95) to [square bottom brace ] (5.9,-0.95);
			    \node [fill=white,inner sep=1pt] at (2.8, -1.05) {$n_3$};
	            \draw [thick] (6.1,-0.2) to [square bottom brace ] (6.7,-0.2);
     		    \node [fill=white,inner sep=1pt] at (6.4, -0.3) {$n_0$};
     		    \tikzset{square bottom brace/.style={ncbar=0.1cm}}
	            \draw [thick] (6.1,-0.45) to [square bottom brace ] (7.5,-0.45);
			    \node [fill=white,inner sep=1pt] at (6.8, -0.55) {$n_1$};
			     \draw [thick] (6.1,-0.7) to [square bottom brace ] (9.1,-0.7);
			    \node [fill=white,inner sep=1pt] at (7.6, -0.8) {$n_2$};		
			    \draw [thick] (9.3,-0.2) to [square bottom brace ] (9.9,-0.2);
     		    \node [fill=white,inner sep=1pt] at (9.6, -0.3) {$n_0$};
	            \draw [thick] (9.3,-0.45) to [square bottom brace ] (10.7,-0.45);
			    \node [fill=white,inner sep=1pt] at (10, -0.55) {$n_1$};			    
			    \draw [thick, blue, dashed] (6.0, 0.6) to (6.0, -1.6);
			    \draw [thick, blue, dashed] (9.2, 0.6) to (9.2, -1.6);
			    \draw [thick, blue, dashed] (10.8, 0.6) to (10.8, -1.6);
			    \node[invisible, blue] at (2.8, -1.4) {Round 1};
			    \node[invisible, blue] at (7.6, -1.4) {Round 2};
			    \node[invisible, blue] at (10, -1.4) {Round 3};
				\end{tikzpicture}
			\end{minipage}
			\caption{Schematic overview of algorithm \failoverstochastic.}
			\label{fig:stoch_alg_illustration}
	\end{figure}
	
		Before describing the algorithm in more detail, an important question that arises is how to use the templates to schedule the future demands. A crucial component in this process are \emph{monotone matchings}, which only match two values if the second is at least as big as the first. 

	\rnote{should be careful that this def works for the convergence part too, or just dont use it later}\mnote{I updated the def, I will double check it's ok}
	\begin{definition}[Monotone matching] Given two sequences  $x_1, \dots, x_n \in \mathbb{R}$ and $y_1, \dots, y_n \in \mathbb{R}$, a \emph{monotone matching} $\pi$ from the $x_t$'s to the $y_t$'s is an injective function from a subset $I \in \{1, \dots, n\}$ to $\{1, \dots, n\}$ such that $x_i \le y_{\pi(i)}$ for all $i \in I$. We say that $x_i$ is \emph{matched} to $y_{\pi(i)}$ if $i \in I$, and $x_i$ is \emph{unmatched} otherwise. 
	\end{definition}
	
	Monotone matchings will allow us to match future demands ($x_i$'s) to the demands that are part of a template ($y_{\pi(i)}$'s) and put the former in the place of the latter (since $x_t \le y_{\pi(i)}$). A surprising result of Rhee and Talagrand~\cite{rhee1993line} is that if the two sequences are sampled i.i.d. from the same distribution, then almost all items can be matched, and moreover such a matching can be found online (see the paper for a more general result where the sequences may come form different distributions). 
	
	\begin{thm}[Monotone Matching Theorem~\cite{rhee1993line}] \label{thm:rheeTal}
		Suppose the random variables $A_1,\ldots,A_n$ and $B_1,\ldots,B_n$ are all sampled independently from a distribution $\mu$. Then there is a constant $\cst$ such that with probability at least $1 - e^{-\cst \cdot \log^{3/2} n}$ there is a monotone matching $\pi$ of the $A_i$'s to the $B_i$'s where at most $\cst \cdot \sqrt{n} \log^{3/4} n$ of the $A_i$'s are unmatched.
%
		Moreover, this matching can be computed even if the sequence $A_1,\ldots,A_n$ is revealed online. 
	\end{thm}

	\subsection{Algorithm}
	
	We are now ready to present the details of the \failoverstochastic algorithm.

	\paragraph{FailoverStochastic:} The algorithm just repeatedly calls the procedure \oneround below, passing to it the number of machines that are still available/unopened (e.g. initially it calls $\oneround(m)$); it does this for $\frac{\log m}{\log 4/3}$ rounds. 
	

	\paragraph{OneRound($\tilde{m}$):}	
	This procedure receives as input the number $\tilde{m}$ of machines that it is allowed to open. It is convenient to rename the demands and use $Y_t$ to denote the $t$th demand seen by \oneround (which are still sampled i.i.d. from $\mu$). Similar to the work of Rhee and Talagrand~\cite{rhee1993line}, this algorithm works in phases: As mentioned earlier, each phase $k$ sees the previous $n_k = 2^k$ items and creates a template based on them, which will then be used to schedule the next $n_k$ items. To create this template, we define the offline problem \minoff of minimizing the number of machines that are required to schedule these $n_k$ items. To solve this problem, we design an approximation algorithm \offlineminmach in \Cref{sec_offline_min_machines} achieving a sublinear approximation guarantee. Specifically, let $\overline{\mach}(x_1,\ldots,x_n)$ be the number of machines that \offlineminmach (with $\e = 1/n_k^{1/6}$) uses to schedule the demands $x_1,\ldots,x_n$. \oneround is then as follows:

\begin{algorithm}[H]
\small
\caption*{\textbf{\oneround:} Given a number of available machines $\tilde{m}$:}
\begin{algorithmic}[1]
	\itemsep=1.5ex
	
	\vspace{1ex}
		\State Assign the first demand  $Y_1$ to an empty edge by itself, opening 2 machines.
		
		\State For phases $k = 0,1,2,\ldots$ \label{algo:step2}
		
		\begin{enumerate}[(a)]
		\item See the first $n_k$ items $Y_1,\ldots,Y_{n_k}$. Run the algorithm \offlineminmach  from Section \ref{sec_offline_min_machines} (with  $\e = 1/n_k^{1/6}$) to find a solution for them that uses $\overline{\mach}(Y_1,\ldots,Y_{n_k})$ machines; let $templ(t)$ denote the pair of machines that $Y_t$ is assigned to. This solution is our template. 
		
		\item If $$\textrm{\#\{already open machines\}} + \underbrace{\overline{\mach}(Y_1,\ldots,Y_{n_k})}_{\textrm{\it machines from template}} + \underbrace{\csta \cdot \sqrt{n_k}\, \log^{3/4} n_k}_{\textrm{\it predicted unmatched demand}} + 2 m^{5/6} \,>\, \tilde{m},$$ then STOP. \label{algo:stepB}
		
		\item Else, open a clique of $\overline{\mach}(Y_1,\ldots,Y_{n_k})$ machines. Upon the arrival of each of the next $n_k$ demands $Y_{n_k + 1}, \ldots, Y_{2n_k}$, assign them to machines based on the template. 
		More precisely, find the Rhee-Talagrand monotone matching $\pi$ guaranteed by \Cref{thm:rheeTal} from the new to the old demands (as the new ones arrive online). Schedule each matched new demand $Y_t$ to the pair of machines that $Y_{\pi(t)}$ occupied in the template, namely the machine pair $templ(\pi(t))$. 
		For each unmatched new demand, schedule it on an edge by itself (opening two more machines for each). If at any point the execution tries to open more than $\tilde{m}$ machines, declare FAIL. 
		\label{algo:stepC}
		\end{enumerate}
\end{algorithmic}
\end{algorithm}

    \subsection{Analysis}
	
	We next discuss the main ideas for the analysis of the algorithm \failoverstochastic, leading to the proof of \Cref{thm_main_stoch}. We assume throughout that $m$ is at least a sufficiently large constant, else the success probability $1 - O(\frac{1}{m})$ trivially holds.\mnote{This is used in Claim \ref{claim:MU} for example. See exactly where to state this assumption, when to recall it, etc.}
	
	We need to develop two important components for the analysis that are done in their own sections. To at least state them, let $\mach(J)$ denote the minimum number of devices needed to assign all demands from set $J$ satisfying the Nominal and Failover constraints. 
	
	\paragraph{First component {\normalfont (\Cref{sec_offline_min_machines})}:}  The first component is the aforementioned algorithm \offlineminmach that is called within \oneround. 
	It relies on a novel configuration LP, \eqref{eq_lpconfigmin}, and a post-processing algorithm to realize a rounded LP solution as a feasible assignment. It has the  following guarantee:
	
		\begin{restatable}{thm}{thmmainoffmin}\label{thm_main_off_min}
			There exists a poly-time algorithm, \offlineminmach, that given $\e \in (0,1)$, finds a solution for \minoff with at most 
			$\big(1 + O(\e)\big) \lpconfigmin + O(\frac{1}{\e^5}) \leq \big(1 + O(\e)\big) \mach + O(\frac{1}{\e^5})$ machines.
		\end{restatable}
		Choosing $\e$ appropriately, we will be able to create a template using at most $\E \, \mach(Y_1, \dots, Y_{n_k})\, + o(n_k)$ devices in expectation for the next $n_k$ arrivals.

	\paragraph{Second component {\normalfont (\Cref{sec:asConverge})}:} Recall from the technical overview that a worrisome aspect of \failoverstochastic is that each call to \oneround does not re-use machines from previous rounds. To show this is not too wasteful, we show that $\E\, \mach(X_1, \ldots, X_T)$ 
	is approximately linear in $T$. We do so by giving a quantitative convergence theorem of $\E\, \mach(X_1, \ldots, X_T)$ to $T \cdot c(\mu)$, where $c(\mu)$ is a constant that characterizes the ``average number of devices needed per demand.'' That is, we show the following:
	
		\begin{restatable}{thm}{thmconverges}\label{thm:converges}
			Let $\mu$ be a distribution supported on $[0,\min\{1, \frac{B}{2}\}]$. Then there exists a scalar $c(\mu)$ 
			such that for every $T \in \mathbb{N}$, we have $$\E\, \mach(X_1,\ldots,X_T) \,\in\, T \cdot c(\mu) \pm O(T^{5/6}),$$
			where $X_1, \ldots, X_T$ are i.i.d. samples from $\mu$.
		\end{restatable}
		Thus splitting the first $2 n_k$ demands into two rounds of $n_k$ demands each costs us only an extra $o(n_k)$ devices.

    \bigskip
	With those two results in hand, the core of the analysis is that \oneround gets good value density, i.e., the ratio of value over number of machines $m$. We use $\E S_0$ to denote the expected value of the size of a demand (which is the same as $\E S_t$ for any $t$).

	
	Specifically, according to \Cref{thm:converges}, there is a scalar $c(\mu)$ such that \OPT is able to fit roughly $\frac{1}{c(\mu)}$ demands per machine. Each such demand gives value roughly $\E S_0$; so the intuition is that the best possible density value/machine should be around $\frac{\E S_0 }{c(\mu)}$. We first make this formal in the next lemma.


	
	\begin{lemma} \label{lemma:stochDensityOPT}
		With probability at least $1 - \frac{2}{m^2}$ we have $$\OPT \le m \cdot \frac{\E S_0}{c(\mu)} + O(m^{5/6}).$$
	\end{lemma}
	
	Crucially, the next lemma says that \oneround almost achieves this density.\mnote{Can we make the term $O(m^{5/6})$ slightly smaller, so that it times $\log m$ is at most $O(m^{5/6})$? Would clean up the bounds a bit}
		
	\begin{lemma} \label{lemma:stochDensity}
		Let $Open$ be the number of machines opened by $\oneround(\tilde{m})$ (which is a random variable). Then with probability at least $1 - \frac{1}{m^2}$, the total value of the demands scheduled by $\oneround(\tilde{m})$ is at least $$\textrm{value of $\oneround(\tilde{m})$} \ge \frac{\E S_0}{c(\mu)} \cdot Open - O(m^{5/6}).$$ 
	\end{lemma}
	
	Given this lemma, we see that the total value of the \failoverstochastic algorithm (which repeatedly calls \oneround) is approximately $\frac{\E S_0}{c(\mu)}$ times the total machines opened during the execution. By showing that the number of machines \failoverstochastic opens is $\approx m$, we then almost match the upper bound on \OPT from  \Cref{lemma:stochDensityOPT}.
		
	
	\begin{lemma} \label{lemma:stochLogCalls}
		With probability $1 - O(\frac{1}{m})$, \failoverstochastic opens at least $m - 5 \cstg \cdot m^{5/6}$ machines (where $\cstg$ is the constant from \Cref{lemma:openCst}).
	\end{lemma}
	
	These lemmas quickly lead to the proof of \Cref{thm_main_stoch}.\mnote{Since it is straightforward, could just give a proof sketch or omit it from the body}
	
	\begin{proof}[Proof of \Cref{thm_main_stoch}]
	Let $L:= \frac{\log m}{\log 4/3}$ denote the number of calls to \oneround that \failoverstochastic makes, and let $val_i$ and $Open_i$ be the value obtained and number of machines opened by the $i$-th call. Employing \Cref{lemma:stochDensity} on these $L$ calls, we have that with probability at least $1 - \frac{L}{m^2}$ the total value of \failoverstochastic is
	\begin{align*}
		\textrm{algo value} ~=~ val_1 + \ldots + val_L ~\ge~ \frac{\E S_0}{c(\mu)} \cdot \sum_{i \le L} Open_i \,-\,O(m^{5/6} \log m).
	\end{align*}
	Moreover, from \Cref{lemma:stochLogCalls}, with probability at least $1-O(\frac{1}{m})$ the total number of machines open $\sum_{i \le L} Open_i$ is at least $m - 5 \cstg \cdot m^{5/6}$, in which case we get 
	\begin{align}
		\textrm{algo value} ~\ge~ m\cdot \frac{\E S_0}{c(\mu)} \,-\,O(m^{5/6} \log m). \label{eq:mainStochAlgVal}
	\end{align}
	Furthermore, from  \Cref{lemma:stochDensityOPT} we have that $\OPT \le m \cdot \frac{\E S_0}{c(\mu)} + O(m^{5/6})$ with probability at least $1 - \frac{2}{m^2}$. So by taking a union bound and combining this with the above lower bound on the algorithm's value, we get that with probability $1 - O(\frac{1}{m})$
	\begin{align*}
		\textrm{algo value} ~\ge~ \OPT \,-\,O(m^{5/6} \log m).
	\end{align*}		
	Since \eqref{eq:mainStochAlgVal} also implies that $\OPT \ge \Omega(m)$, the previous bound is at least $\OPT \,-\,O(\opt^{5/6} \log \opt)$. This concludes the proof of \Cref{thm_main_stoch}.
	\end{proof}	

	We conclude this section by proving the lower bound on the value density of \oneround from \Cref{lemma:stochDensity}. We defer the proofs of \Cref{lemma:stochDensityOPT} and \ref{lemma:stochLogCalls} to \Cref{app:stoch_proofs}.


	\subsubsection{Proof of \Cref{lemma:stochDensity}} 

	First, we control in high-probability the number of phases that $\oneround(\tilde{m})$ executes before stopping or failing; this will be important to avoid dependencies on the total number of demands $n$ in the instance, which can be arbitrarily bigger than the scale of the effective instance. 

	\begin{claim} \label{claim:numPhases}
		With probability at least $1 - \frac{1}{m^3}$ the algorithm \oneround performs at most 
	\begin{align}
	\bar{k} := \log \bigg(\frac{\tilde{m}}{c(\mu)} + O(\tilde{m}^{5/6}) + 3 \log^{\frac{3}{2}} m 
	\bigg) \label{eq:barK}
	\end{align} \knote{I believe the last term needs to become $3 \log^{\frac{3}{2}} m$ after the changes in the lemma in the appendix}
	phases.
	\end{claim}

	\begin{proof}
	Recall that the demands sizes $Y_1,Y_2,\ldots$ that \oneround sees are still i.i.d. samples from the original distribution $\mu$. In \Cref{lemma:numScheduled} (with $m=\tilde{m}$ and $\delta=\frac{1}{m^3}$) in the appendix we show that with probability at least $1 - \frac{1}{m^3}$ \oneround can schedule at most $\frac{\tilde{m}}{c(\mu)} + O(\tilde{m}^{5/6}) + 3 \log^{\frac{3}{2}} m 
	$\knote{Change last term based on the new lemma} many of these demands; for some intuition, \Cref{lemma:stochDensityOPT} indicates that even $\opt$ cannot schedule more than roughly these many demands. Since this quantity is exactly $n_{\bar{k}}$, \oneround cannot complete phase $\bar{k}$ (there are $2 n_{\bar{k}}$ demands by the end of it) and the claim holds. 
	\end{proof}	
	
	Next, we need to bound how many machines are opened by \oneround, which in particular affects the probability of it failing. For a phase $k$, let $M_k := \overline{\mach}(Y_1,\ldots,Y_{n_k})$ denote the number of machines in the template solution, and let $U_k$ be the number of additional machines that had to be open to accommodate the unmatched demands among $Y_{n_k +1},\ldots,Y_{2n_k}$, namely twice the number of unmatched items. Notice that these quantities are well defined even for phases that the algorithm did not execute. The quantity $M_k + U_k$ is then the number machines that the algorithm \oneround opens in phase $k$ (if it executes it). We have the following bounds for the number of machines open, at least for a phase $k$ where the number of items $n_k$ is sufficiently large (but still sublinear in $m$).
	
	\begin{claim} \label{claim:MU}
	  Let $k_0 := (\frac{2}{\cstc} \log m)^{2/3}$ \knote{not that it makes much difference but the 2 should be 3 after the changes I think (due to 3 in $m^3$)} for a sufficiently small constant $\cstc$. Then there is a constant $\csta$ such that:\vspace{-9pt}
	  
	  \begin{enumerate}
	  	\item For $k \ge k_0$, we have $M_k \,\in\, n_k \cdot c(\mu) \,\pm\, \csta \cdot n_k^{5/6}$ with probability $\ge 1 - \frac{1}{m^3}$ \vspace{-5pt}
	  	
	  	\item For $k \ge k_0$, we have $U_k \le  \csta \cdot \sqrt{n_k}\,\log^{3/4} n_k$ with probability $\ge 1 - \frac{1}{m^3}$ \vspace{-5pt}
	  	
 		\item $n_{k_0} \le m^{5/6}$.
		\end{enumerate}
	\end{claim}
	
	\begin{proof}
	 Consider a phase $k \ge k_0$. Since the demand sizes $Y_1,\ldots,Y_{2n_k}$ seen in this phase are i.i.d. samples from the original distribution $\mu$, we can bound the minimum number of machines  $\mach(Y_1,\ldots,Y_{n_k})$ (using \Cref{cor:convergesWHP} in \Cref{app:proofStochDensityOPT} with $\lambda = n_k^{1/3}$)  
	 \begin{align*}
	 	\mach(Y_1,\ldots,Y_{n_k}) \,&\in\, n_k \cdot c(\mu) \pm O(n_k^{5/6})
	 \end{align*}
	 with probability at least $1 - 2e^{-\frac{n_k^{2/3}}{2}}$. Moreover, employing the guarantee of the algorithm \offlineminmach 
	 used to build the template (\Cref{thm_main_off_min} with $\e = 1/n_k^{1/6}$), we get 
	 %
	 \begin{align*}
	 	M_k \,&\in\, n_k \cdot c(\mu) \pm \csta \cdot n_k^{5/6}
	 \end{align*}
	 with probability at least $1 - 2e^{-\frac{n_k^{2/3}}{2}}$ for some constant $\csta$. But since $n_k = 2^k \ge 2^{k_0}$, a quick calculation shows that this probability is at least $1 - \frac{1}{m^3}$, proving the first item of the claim.  	 

	 To control $U_k$, we can use the Monotone Matching Theorem (\Cref{thm:rheeTal}) with the first sequence of sizes being the demands from the template, i.e., $(B_1,\ldots,B_{n_k}) = (Y_1,\ldots,Y_{n_k})$, and the second one being the demands that we attempted to match to them, namely $(A_1,\ldots,A_{n_k}) = (Y_{n_k+1}, \ldots,Y_{2n_k})$ to obtain that the number of unmatched demands it at most $\cst \cdot \sqrt{n_k}\,\log^{3/4} n_k$ with probability at least $1- e^{-\cst \cdot \log^{3/2} n_k}$, and hence with this probability
	 \begin{align*}
	 	U_k \,\le\, 2 \cst \cdot \sqrt{n_k}\,\log^{3/4} n_k. 
	 \end{align*}
	Again because $k \ge k_0$, we get that this probability is at least $1- \frac{1}{m^3}$, proving Item 2 of the claim (by taking $\csta \ge 2\cst$ we can just replace the latter by the former). 
	
	The last item $n_{k_0} \le m^{5/6}$ of the claim can be directly verified using the fact that we assumed $m$ is at least a sufficiently large constant. 
	\end{proof}
	 
	Recall that \oneround only fails when the number of machines $M_k + U_k$ actually opened in a phase is bigger than it ``predicted'' in Line \ref{algo:step2}.\ref{algo:stepB}, and this prediction is exactly $M_k$ plus the upper bound $U_k$ from \Cref{claim:MU} plus a slack of $2 m^{5/6}$. By considering all phases, it is now easy to upper bound the probability that \oneround fails ($\bar{k}$ is defined in \eqref{eq:barK}). 
	
	\begin{claim} \label{claim:failPr}
	The probability that \oneround fails is at most $\frac{\bar{k} + 1}{m^3}$.
	\end{claim}
	
	\begin{proof}
Fix any phase $k$, and we claim that the probability that \oneround fails on this phase is at most $\frac{1}{m^3}$. If \oneround fails on phase $k$, then it did not STOP in Line \ref{algo:step2}.\ref{algo:stepB}, so $$\textrm{\#\,[machines open before phase $k$]} + \overline{\mach}(Y_1,\ldots,Y_{n_k}) + \csta \cdot \sqrt{n_k}\, \log^{3/4} n_k + 2m^{5/6} ~\le~ \tilde{m},$$ but it ran out of machines during phase $k$, namely  $$\textrm{\#\,[machines open before phase $k$]} + (M_k + U_k) > \tilde{m}.$$ Since $M_k = \overline{\mach}(Y_1,\ldots,Y_{n_k})$, these observations imply that $U_k > \csta \cdot \sqrt{n_k}\, \log^{3/4} n_k + 2m^{5/6}$. This is impossible if $n_k \le m^{5/6}$, because the number of machines $U_k$  opened for the unmatched demands is at most twice the number $n_k$ of demands considered for the matching. So we must have $n_k > m^{5/6}$ (and so from \Cref{claim:MU} $k \ge k_0$) and at least $U_k > \csta \cdot \sqrt{n_k}\, \log^{3/4} n_k$; but again by \Cref{claim:MU} the latter happens with probability at most $\frac{1}{m^3}$. Thus, the probability that \oneround fails on phase $k$ is at most $\frac{1}{m^3}$.

	Moreover, by \Cref{claim:numPhases}, with probability at least $1- \frac{1}{m^3}$ \oneround has at most $\bar{k}$ phases. Then taking a union bound, we see that the event that \oneround has at most $\bar{k}$ phases and in all of them it does not fail holds with probability at least $1 - \frac{\bar{k} + 1}{m^3}$; in particular, with at least this much probability the algorithm does not fail in its execution, which proves the claim.
	\end{proof}
	 
	We now finally lower bound the value that \oneround gets. Let $\tau$ be the (random) index of the last phase attempted by \oneround, namely where Line \ref{algo:step2}.\ref{algo:stepC} is executed. As long as it does not fail on the last phase $\tau$ (which by the previous claim happens with probability at least $1 - \frac{\bar{k}+1}{m^3}$) \oneround gets the value of all items up until this phase, that is
	\begin{align}
	\textrm{value of \oneround} \,\ge\, Y_1 + \ldots + Y_{2n_\tau} \,\ge\, Y_1 + \ldots + Y_{2n_{\blue{\min\{\tau,\bar{k}\}}}}.  \label{eq:valOneRound}
	\end{align}
		Recall that the $Y_i$'s are independent and each has mean $\E S_0$. Then employing the Chernoff bound (\Cref{lemma:chernoff}) with $\lambda = \sqrt{n_{\bar{k}} \log (m^3 \cdot n_{\bar{k}})}$, for any fixed $t \le n_{\bar{k}}$ we have that 
	 \begin{align*}
	 Y_1 + \ldots + Y_t \ge t \cdot \E S_0 - \sqrt{n_{\bar{k}} \log (m^3 \cdot n_{\bar{k}})} ~~~~~~\textrm{with probability at least $1 - \frac{1}{m^3 \cdot n_{\bar{k}}}$}.
	 \end{align*}
	 Then taking a union bound over \eqref{eq:valOneRound}, the previous displayed inequality for all $t \le n_{\bar{k}}$, and over the event that \oneround has at most $\bar{k}$ phases (which holds with probability at least $1- \frac{1}{m^3}$) we get that
	\begin{align}
	\textrm{value of \oneround} \,&\ge\, 2n_{\min\{\tau,\bar{k}\}} \cdot \E S_0 - \sqrt{n_{\bar{k}} \log (m^3 \cdot n_{\bar{k}})}  \notag\\
	&= 2n_\tau \cdot \E S_0 - \sqrt{n_{\bar{k}} \log (m^3 \cdot n_{\bar{k}})}\notag\\
	&\ge 2n_\tau \cdot \E S_0 - O(m^{5/6}) ~~~~~~~~~~~~\textrm{with probability $\ge 1 - \frac{\bar{k} + 3}{m^3}$}.   \label{eq:valOneRound2}
	\end{align}
	 
	To conclude the proof of \Cref{lemma:stochDensity} we just need to relate this quantity to the number of machines opened by \oneround. Let $Open_\ell$ be the number of machines opened until (including) phase $\ell$, and recall that $Open$ is the number of machines opened over all phases. Since the number of machines opened on phase $k$ is $M_k + U_k$ (plus two machines for the first demand $Y_1$), we have 
	\begin{align}
	Open_\ell = 2 + (M_1 + U_1) + \ldots + (M_\ell + U_\ell)  \label{eq:openEll}
	\end{align}

	To upper bound the right-hand side, for the phases $k < k_0$ we just use the fact that $M_k + U_k \le 2n_k + 2n_k = 4n_k$, since both in the template and for the unmatched demands we never open more than 2 machines per demand considered (and $n_k$ demands are considered in each part). For each phase $k = k_0, \ldots, \bar{k}$ we can use \Cref{claim:MU} to upper bound $M_k + U_k$ with probability at least $1 - \frac{2}{m^3}$ by
	\begin{align*}
	M_k + U_k &\le n_k \cdot c(\mu) + \csta \cdot n_k^{5/6} + \csta \cdot \sqrt{n_k}\,\log^{3/4} n_k\\
	&\le n_k \cdot c(\mu) + \cstd \cdot n_k^{5/6}
	\end{align*}
	for some constant $\cstd$. Together these bounds give that with probability at least $1 - \frac{2\ell}{m^3}$
	\begin{align*}
	Open_\ell &\le 2 + \sum_{k < k_0} 4 n_k + \sum_{k = k_0}^{\ell} \bigg( n_k \cdot c(\mu) + \cstd \cdot n_k^{5/6} \bigg).
	\end{align*}	
	To further upper bound the first summation on the right-hand side, because of the exponential relationship $n_k = 2^k$, we have $\sum_{k < k_0} 4n_k \le 8 n_{k_0-1} \le O(m^{5/6})$, the last inequality coming from \Cref{claim:MU}; for the second summation, we analogously have $\sum_{k=k_0}^{\ell} n_k \le 2 n_\ell$ and $\sum_{k = k_0}^{\ell} n_k^{5/6} \le O(n_{\ell}^{5/6})$. Therefore,
	\begin{align}
	Open_\ell \,\le\, 2 n_\ell \cdot c(\mu) + O(n_{\ell}^{5/6}) + O(m^{5/6})~~~~~\textrm{with probability at least $1 - \frac{2\ell}{m^3}$}. \label{eq:openEll2}
	\end{align}
	Finally, since by \Cref{claim:numPhases} the number of phases $\tau$ performed by \oneround is at most $\bar{k}$ with probability at least $1 - \frac{1}{m^3}$, the total number of machines open can be upper bounded  
	\begin{align*}
		Open \,\le\, Open_{\min\{\tau,\bar{k}\}} \,\le\, 2 n_{\tau} \cdot c(\mu) + O(n_{\bar{k}}^{5/6}) + O(m^{5/6}) \,\le\, 2 n_{\tau}\cdot c(\mu) + O(m^{5/6})
	\end{align*}
	with probability at least $1 - \frac{2 \bar{k} + 1}{m^3}$. 
	
		Finally, taking a union bound to combine this inequality with \eqref{eq:valOneRound2}, we get that 
	\begin{align*}
		\textrm{value of \oneround} &\,\ge\, \frac{\E S_0}{c(\mu)} \cdot Open - O(m^{5/6})
	\end{align*}
	with probability at least $1 - \frac{3 \bar{k} + 4}{m^3}$. Since $m$ is at least a sufficiently large constant, we have $m \ge 3\bar{k} + 4$, and the bound from the displayed inequality holds with probability at least $1 - \frac{1}{m^2}$. This finally concludes the proof of \Cref{lemma:stochDensity}.

\ifx \hasmain \undefined

\usepackage{fullpage}
\usepackage[pdftitle={Tile},
            pdfauthor={Author},
            pdfkeywords={Keywords}]{hyperref}

\input{preamble}

\begin{document}

\fi
	
\section{Offline Machine Minimization}\label{sec_offline_min_machines}

	In this section we consider the aforementioned  (offline) minimization version of \problem, which we call \minoff: Given a failover capacity $B \geq 1$ and a collection of demands such that demand $j$ has size $s_j \in [0, \min\{1, \frac{B}{2}\}]$, we need to assign \emph{all demands} to pairs of machines while satisfying the Nominal and Failover constraints, and the goal is to minimize the number of machines used. As before, we use $\mach = \mach(s_1,\ldots,s_n)$ to denote the cost of (i.e. number of machines in) the optimal solution.


The main result of this section (\Cref{thm_main_off_min}, restated) is an efficient algorithm with a sublinear additive regret for this problem (when $\e$ is set appropriately). We remark that a sublinear regret (compared to, say, a constant approximation) is necessary due to its use in \Cref{sec:stoch}. In fact, the algorithm compares against the stronger optimum of an LP relaxation for the problem (denoted by \eqref{eq_lpconfigmin}, and defined below), which will be crucially used in \Cref{sec:asConverge}. We let $\lpconfigmin$ denote the optimal value of this LP.\mnote{Maybe say a bit more why it is important there}

\thmmainoffmin*
	As hinted above, our algorithm is based on converting a solution of a configuration LP into a good assignment of demands to pairs of machines. But crucially, while the configuration of each machine controls the total size of demands serviced by it, it has no information how these demands are distributed over the ``edges'' incident to the machine, which is important for adequately handling the Failover constraints. The post-processing of the LP solution is the one in charge of creating a feasible (and low-cost) assignment from this limited control offered by the LP. \mnote{Maybe move (a version of this) to the intro?}
		
	 

\subsection{Configuration LP}

Consider an assignment of the demands into some number of machines. We can view the collection of demands assigned to (the edges incident to) a given machine as a configuration. Precisely, we define a \emph{configuration} $C$ to be a subset of the demands such that $\sum_{s \in C} s \leq 1$ and $\sum_{s \in C} s + \max_{s \in C} s \leq B$. Note that the first constraint is exactly the Nominal constraint, while the second is a relaxation of the Failover constraint, because the most-loaded edge incident on some machine can be larger than the single largest demand assigned to that machine. Thus, our notion of configuration does not take in to account how the demands are assigned to the respective edges incident on each machine.

To define our configuration LP, we suppose the input collection of demands is partitioned into $T$ \emph{demand types} such that type $t$ consists of $n_t$-many demands each with size $s_t$. Thus each configuration $C$ can be represented by a number $n_t(C) \in \mathbb{N}$ of demands for each type $t$ such that $\sum_t n_t(C) \cdot s_t \leq 1$ and $\sum_t n_t(C) \cdot s_t + \max_{t \mid\, n_t(C) > 0} s_t \leq B$. We are ready to define our configuration LP:

    \begin{equation}\tag{$\lpconfigmin$}\label{eq_lpconfigmin}
		\begin{array}{rrll}
			\min & \sum_C x_C\\
			s.t. & \sum_C n_t(C) \cdot x_C & \geq 2 n_t &\quad \forall t\\
			& x & \geq 0
		\end{array} 
	\end{equation}

Note that the definition of \eqref{eq_lpconfigmin} depends on how the demands are partitioned into types. We show in \Cref{sec:consistencyLP} that the optimal value of \eqref{eq_lpconfigmin} does not depend on the particular type partition. Thus, throughout the analysis, we will use whichever type partition is convenient (unless a particular one is specified).

It is immediate that \eqref{eq_lpconfigmin} is a relaxation of \minoff by taking the natural setting of the $x$-variables defined by a feasible assignment to machines: just let $x_C$ be the number of machines whose collection of demand sizes assigned to its edges are exactly those in $C$. In particular, we have that $\lpconfigmin \le \mach$.



Although \eqref{eq_lpconfigmin} has exponentially many variables in general, we can approximately solve it via column generation similar to the standard bin packing configuration LP~\cite{karpKarmarkar,rothvossConfigLP} (see \Cref{sec:solve}). 

\begin{lemma}\label{lem_lpconfigmin_solve}
	We can find in poly-time an extreme point solution of $\eqref{eq_lpconfigmin}$ with objective value at most $\lpconfigmin + 1$.
\end{lemma}

Further, observe that \eqref{eq_lpconfigmin} only has $T$ non-trivial constraints, so by the standard rank argument (see for example Lemma 2.1.3 of~\cite{raviIterativeMethods}) any extreme point solution of \eqref{eq_lpconfigmin} has at most $T$ non-zero variables. 
Thus, the next lemma follows immediately by rounding up all the fractional variable of an extreme point solution.

\begin{lemma}\label{lem_lpconfigmin_round}
	Given an extreme point of \eqref{eq_lpconfigmin} with objective value $\textrm{Val}$, rounding up all fractional variables to the next largest integer gives an integral solution to \eqref{eq_lpconfigmin} with objective value at most $\textrm{Val} + T$.
\end{lemma}

To summarize this section, we can efficiently obtain a collection of configurations, each corresponding to a machine, that ``covers'' all the demands. However, these configurations do not specify how to actually assign the demands to the edges incident on the corresponding machine. This is the goal of the next section.

\subsection{Matching configurations}

	We say that a collection $\C$ of configurations is feasible if it comes from an integer solution for \eqref{eq_lpconfigmin}, i.e. setting $x_C$ to be the number of times $C$ appears in $\C$ gives a feasible solution for \eqref{eq_lpconfigmin}. Our goal in this section is to realize such collection by actually assigning demands to edges. The main challenge is satisfying the actual Failover constraints.
	
	For simplicity assume $\sum_{C \in \C} n_t(C) = 2 n_t$ for all types $t$, i.e. each demand appears on exactly 2 configurations (drop from the configurations what is extra). We can think of $\C$ (with, say, $N$ configurations) as a graph on $N$ nodes/machines, where node/machine $C \in \C$ has $n_t(C)$ ``slots'' for demands of type $t$. While this gives the right number of slots $2n_t$ to accommodate the demands of each type $t$, we still need to specify to which \emph{edge} (pair of machines) each of the $n_t$ demands of type $t$ is assigned in a way that satisfies the Nominal and Failover constraints. (We can alternatively see this as a graph realization problem: each node $C$ as having a requirement $n_t(C)$ of ``edges of type $t$'' (which we call its \emph{$t$-degree}) and we want to create edges of different types (i.e., assignment of demands to pairs of nodes) to satisfy these requirements while also satisfying the Nominal and Failover constraints.)
	
To see the challenge, consider a fixed node/configuration $C$. Regardless of how we assign demands to edges (as long as it is consistent with the slots of the configurations), the Nominal constraint of $C$ is satisfied: it will receive total size $\sum_t n_t(C) \cdot s_t = \sum_{s \in C} s$, which is at most $1$ by definition of a configuration. This is not the case for the Failover constraint. This is again because the definition of configuration only gives us the relaxed version of the Failover constraint $\sum_{s \in C} s + \blue{\max_{s \in C} s} \leq B$,\mnote{Keep this blue!} In particular, the blue term only considers the largest demand assigned to machine $C$ instead of the most-loaded edge incident to $C$. However, these two quantities are the same \emph{if we are able to assign at most one demand per edge}. (In the graph realization perspective, it means that it suffices to construct a \emph{simple} graph with the desired $t$-degrees.) But it is not clear that such an assignment should even exist, let alone be found efficiently. 

The main result of this section is that -- by opening slightly more machines -- we can find such an assignment that realizes any given collection of configurations satisfying both the Nominal  and Failover constraints.

\begin{thm}\label{thm_config_match}
	Consider an instance of \minoff with $T$ demand types. Given a collection $\C$ of $N$ configurations that is feasible for \eqref{eq_lpconfigmin}, we can find in poly-time a feasible solution for \minoff that uses at most $N + O(DT)$ machines, where $D$ is the maximum number of demands in any configuration in $\C$.
\end{thm}

	For that, we will need the following subroutine to assign some demands outside of their respective configurations.

\begin{lemma}\label{lem_matching_unit}
	There is a poly-time algorithm for \minoff that uses at most $8 \cdot S + 2$ machines, where $S$ is the sum of the size of the demands in the instance. 
\end{lemma}

\begin{proof}
	Our algorithm will only open edges -- that is, we will open machines in pairs and will only assign demands to the edges of the paired machines. Our algorithm is the following: Consider the demands in any order. Assign each demand to already-opened edge as long as the Nominal and Failover constraints remain satisfied. Else open a new edge and assign the demand there.
	
	It is clear that the algorithm is efficient and satisfies the constraints. We claim this algorithm opens at most $8 \cdot S + 2$ machines (i.e. $4 \cdot S + 1$ edges). To see this, note that every edge except at most one has load at least $\frac{1}{4}$. If not, then consider the first time that there are two open edges with load less than $\frac{1}{4}$. It must be the case that the last demand $s_j$ the algorithm considered up to this point had size less than $\frac{1}{4}$, but the algorithm decided to open a new edge for this demand rather than assign it to a previous edge $(u,v)$ that already had load less than $\frac{1}{4}$. However, assigning demand $s_j$ to the edge $(u,v)$ is feasible; the left-hand side of the Failover constraint for machine $u$, say, would be at most $\frac{1}{2}$ (total size assigned to machine $u$) plus $\frac{1}{2}$ (total size assigned to the edge $(u,v)$, the only one incident to $u$), which is at most the Failover capacity $B \ge 1$.) This contradicts the definition of the algorithm.
\end{proof}

The algorithm guaranteed by \Cref{thm_config_match} is the following. In order to simplify the notation, as before we assume without loss of generality that $\C$ has $\sum_{C \in \C} n_t(C) = 2n_t$ for all types $t$. 

\begin{algorithm}[h]
\small
\caption*{\textbf{\matchalg:} Given a collection $\C$ of $N$ configurations:}
\begin{algorithmic}[1]
	\itemsep=1.5ex
	
	\vspace{1ex}
	\State Open $N$ machines -- one corresponding to each configuration in $\C$.
	\State Consider demand types in arbitrary order $t = 1, \dots, T$.
	\State When considering demand type $t$, partition the collection $\C$ into two collections $\cL_t$ and $\cR_t$ such that their total $t$-degrees $\sum_{C \in \cL_t} n_t(C)$ and  $\sum_{C \in \cR_t} n_t(C)$ differ by at most $D_t$, where $D_t$ is the maximum number of type $t$ demands in any configuration. (This can be achieved, e.g., by initializing $\cL_t, \cR_t = \emptyset$, and adding configurations one-by-one to the set with minimum total $t$-degree.)
	\State Given this partition, as long as there exists a configuration $C \in \cL_t$ that is currently assigned less than $n_t(C)$ demands of type $t$, we pick such a configuration and assign a demand of type $t$ to an arbitrary edge $(C, C')$ (for $C' \in \cR_t$) that has not yet been assigned a demand of any type and such that $C'$ is currently assigned less than $n_t(C')$ demands of type $t$. If no such edge exists, then we stop and move on to the next demand type.\label{step_one_per_edge}
	\State Once we are done considering all demand types, assign all the currently unassigned demands to new machines using \Cref{lem_matching_unit}.\label{step_unassigned}
\end{algorithmic}
\end{algorithm}

\begin{proof}[Proof of \Cref{thm_config_match}]
	It is clear that \matchalg runs in polynomial time, and assigns all demands to edges. Further, this assignment satisfies both the Nominal and Failover constraints, because we assign at most one demand per edge in Step \ref{step_one_per_edge} (see discussion in the beginning of this section) \mnote{More detail? Or given the discussion above this is ok?}, and Step \ref{step_unassigned} guarantees a feasible assignment for the remaining demands.
	
	It remains to show that it opens $N + O(DT)$ machines. In particular, by \Cref{lem_matching_unit} it suffices to show that the total size of all unassigned demands that reach Step \ref{step_unassigned} is $O(DT)$. When considering demand type $t$, there are two possibilities:
	
	\paragraph{Case 1:} Step \ref{step_one_per_edge} assigns $n_t(C)$ type $t$ demands to each $C \in \cL_t$. In this case it assigns $\sum_{C \in \cL_t} n_t(C)$ type $t$ demands to edges between $\cL_t$ and $\cR_t$, while the total number of type $t$ demands is 
		\begin{align}
			n_t = \frac{1}{2} \bigg(\sum_{C \in \cL_t} n_t(C) + \sum_{C \in \cR_t} n_t(C) \bigg) \le \sum_{C \in \cL_t} n_t(C) + \frac{D_t}{2}, \label{eq:degreeDisc}
		\end{align}
		where the inequality uses the fact that the $t$-degree of $\cR_t$ is at most that of $\cL_t$ plus $D_t$. Thus, at most $\frac{D_t}{2}$ demands of type $t$ remain unassigned and reach Step \ref{step_unassigned}. The total size of these demands it at most $\frac{1}{2}$, since $D_t$ demands of type $t$ are in a valid configuration. Hence the total size of the unassigned demands of all types is at most $\frac{T}{2} \le O(DT)$.
		
	\paragraph{Case 2:} Step \ref{step_one_per_edge} fails to assign $n_t(\bar{C})$ to a configuration $\bar{C} \in \cL_t$. In this case, for each $C' \in \cR_t$, either the edge $(\bar{C},C')$ is already assigned some demand (call such $C'$ \emph{blocked}) or $C'$ has already been assigned $n_t(C')$ demands of type $t$. But there are at most $D$ blocked $C'$'s, since the configuration $\bar{C}$ has at most $D$ slots to receive demands. Thus the total number of type-$t$ demands assigned is at least 
	\begin{align*}
		\sum_{C' \in \cR_t \setminus \textrm{blocked}} n_t(C') \,\ge\, \sum_{C' \in \cR_t} n_t(C') \,-\, D \cdot \max_{C' \in \textrm{blocked}} n_t(C') \,\ge\, \sum_{C' \in \cR_t} n_t(C') \,-\, D \cdot D_t. 
	\end{align*}
	Moreover, exchanging the roles of $\cL_t$ and $\cR_t$ in the argument from \eqref{eq:degreeDisc} we get that $\sum_{C' \in \cR_t} n_t(C') \ge n_t - \frac{D_t}{2}$, and thus at least $n_t - D \cdot D_t - \frac{D_t}{2}$ demands of type $t$ are assigned by Step \ref{step_one_per_edge}. Thus at most $O(D \cdot D_t)$ demands (hence total size $O(D)$) of this type remain unassigned and reach Step \ref{step_unassigned}. This a total size of $O(DT)$, over all demand types, that reach the latter step, as desired.  
\end{proof}

We summarize the main results of this section and the previous with the next theorem: By approximately solving \eqref{eq_lpconfigmin} (\Cref{lem_lpconfigmin_solve}), rounding the solution (\Cref{lem_lpconfigmin_round}), and using the above algorithm to obtain an assignment of demands to edges (\Cref{thm_config_match}), we obtain the following. 

\begin{thm}\label{thm_lpconfig_main}
	Consider an instance of \minoff that has most $T$ demands types and where each configuration has at most $D$ demands. Then there is a poly-time algorithm that finds a feasible solution that uses at most $\lpconfigmin + O(DT)$ machines.
\end{thm}


To use this procedure for obtaining our main result, \Cref{thm_main_off_min}, we need to modify the input instance to make $D$ and $T$ small enough, which is the goal of the next section. 


\subsection{Reducing the number of types and demands in a configuration} \label{sec_small_large_jobs}

	Given an arbitrary set of demands $J$, we will convert this into another set of demands $\tilde{J}$ with small parameters $T$ and $D$ such that: 1) The optimal LP value for $J$ and $\tilde{J}$ are similar; 2) We can convert an assignment of the demands $\tilde{J}$ into an assignment of the original demands $J$ without using many extra machines. 
	
	Given a parameter $\e \in (0,1)$, the instance $\tilde{J}$ is constructed as follows. First, partition $J = S \cup M$ into small ($s_j < \e^2$) and medium demands ($s_j \geq \e^2$), respectively. To reduce the number of types of medium demands we apply \emph{linear grouping}, a transformation used in the context of Bin Packing~\cite{karpKarmarkar}: Let $n := \lvert M \rvert$. Partition $M$ into $\frac{1}{\e^3}$ groups, consisting of the $\e^3 n$ largest demands, the next $\e^3 n$ largest demands, and so on. Note every group has size exactly $\e^3 n$ except possibly the last group, corresponding to the smallest demands. At this point, let $L$ be the first group, corresponding to the largest demands. We can now partition $J = S \cup (M \setminus L) \cup L$. For the remaining groups of demands in $M \setminus L$, let $\tilde{M}$ denote the modified set of demands, where we round the size of each demand in $M \setminus L$ up to the largest size in its group. (The demands $L$ will not be part of the final instance $\tilde{J}$ and have to be handled separately.)
	
	For the small demands $S$, we want to both reduce the number of types but also ensure that none of them are too small (in order to limit the max number of demands in a configuration). Simply rounding all of them to the threshold $\e^2$ may increase the LP value too much, and rounding down to $0$ is effectively ignoring these demands, which makes it difficult to produce an assignment for the original instance $J$. So instead the idea is to group the small items $S$ into ``blocks'' of size exactly $\e$. Since no set of small items may add to exactly this size, we actually just create the appropriate amount of ``blocks'': let $\tilde{S}$ be a collection of $\lceil \frac{1}{\e} \sum_{s \in S} s \rceil$ demands of size $\e$.
	
	Then the transformed instance is given by the modified medium and small items, that is, $\tilde{J} = \tilde{S} \cup \tilde{M}$. The parameters $T$ and $D$ are indeed controlled: there are $T = \frac{1}{\e^3} + 1 = O(\frac{1}{\e^3})$ types, and since all demands have size at least $\e^2$ the max number of demands in a configuration is $D = \frac{1}{\e^2}$. (Also note that the large demands $L$, which are treated separately, have sizes at least $\e^2$ and there are at most $\e^3 \lvert M \rvert$ of them.) The next lemma states that our modifications did not increase $\lpconfigmin$ by much. For any set of demands $J$, we let $\lpconfigmin(J)$ be the optimal LP value for this set of demands (recall that this is well-defined regardless of the partition of demands into types \Cref{sec:consistencyLP}).\mnote{Maybe we can defer the proof to the appendix, and just give a sketch (or nothing :))}\rnote{we could deter to appendix to save space if you want - i think people would believe us if we just stated the lemma}

\begin{lemma}\label{lem_small_med_lp}
	For any set of demands $J$ and $\e \in (0,1)$, let the modified set of demands $\tilde{J} = \tilde{S} \cup \tilde{M}$ be defined as above.
	Then $$\lpconfigmin(\tilde{J}) \leq \big(1 + O(\e)\big)  \lpconfigmin(J) + O(1).$$
\end{lemma}
\begin{proof}
	We first claim that $\lpconfigmin(S \cup \tilde{M}) \leq  \lpconfigmin(J)$. To see this, recall that $\tilde{M}$ is obtained from $M$ by excluding $L$ (the group of the largest $\e^3 n$ demands) and rounding up all remaining groups. Thus, for every group $G$ in $M$ except $L$, there exists a next-larger group $G'$ in $M$ with at least as many demands and such that every demand in $G'$ has size greater than or equal to every demand in $G$. This still holds after rounding up all demand sizes in $G$ to largest demand in $G$, which is how we obtain $\tilde{M}$. Thus, we can monotonely match every item in $S \cup \tilde{M}$ to items in $S \cup M = J$, and with this it is not hard to see that $\lpconfigmin(S \cup \tilde{M}) \leq \lpconfigmin(J)$.
	
	Given this inequality, it suffices to show $\lpconfigmin(\tilde{S} \cup \tilde{M}) \leq \big(1 + O(\e)\big) \lpconfigmin(S \cup \tilde{M}) + O(1)$. We will use an optimal solution for $\lpconfigmin(S \cup \tilde{M})$ to construct a feasible solution to $\lpconfigmin(\tilde{S} \cup \tilde{M})$ without increasing the objective value by much. It is convenient to consider the type partition of $S \cup \tilde{M}$ where every small job is its own type, and there is one type for each group in $\tilde{M}$. Similarly, for $\tilde{S} \cup \tilde{M}$, we consider the type partition with one demand type for all demands in $\tilde{S}$ and one for each group in $\tilde{M}$.
	
	To achieve this, we map each configuration $C$ of demands $S \cup \tilde{M}$ to a configuration $C'$ using demands $\tilde{S} \cup \tilde{M}$ as follows: Starting from $C$, we keep all of its $\tilde{M}$ demands. For the $S$ (small) demands, let $k$ be an integer such that the total size of all small demands in the configuration is in the interval $[k \cdot \e, (k+1) \cdot \e)$. We arbitrarily remove small demands from this configuration until the total size of the remaining small demands lies in the interval $[k \cdot \e - \e^2, k \cdot \e]$. In doing so, we remove at most $\epsilon + \epsilon^2$ units of small demands (using the fact that every small demand has size at most $\epsilon^2$.) Then we replace the $S$ demands in this configuration $C$ with $\max(0, k-1)$ blocks of small demands (i.e. demands in $\tilde{S}$), obtaining a configuration $C'$ of demands  $\tilde{S} \cup \tilde{M}$. 
	
	\begin{prop}
		The configurations $C'$ created by the above procedure are valid configuration.
	\end{prop}
	\begin{proof}
		Let $C$ be the input configuration of demands $S \cup \tilde{M}$ and $C'$ the resulting configuration of demands $\tilde{S} \cup \tilde{M}$ (and throughout, let $k$ be the one used in the transformation). We need to show that $\sum_{s \in C'} s \leq 1$ and $\sum_{s \in C'} s + \max_{s \in C'} s \leq B$. For the first inequality, both $C$ and $C'$ have the same $\tilde{M}$ demands and the total size of $S$ demands in $C$ is at least $k \cdot \e$ and the total size of $\tilde{S}$ demands in $C$ is $\max\{0, (k-1) \cdot \e\}$; thus $\sum_{s \in C'} s \le \sum_{s \in C} s \le 1$, the last step following from the validity of~$C$.  
		
		It remains to show the second inequality. There are two cases to consider:
		
		\begin{enumerate}
			\item If $k \leq 1$ then $C'$ contains no blocks of small demands but only the $\tilde{M}$ demands that are also present in $C$. Then we have
			\begin{align}
			\sum_{s \in C'} s + \max_{s \in C'} s \,\le\, \sum_{s \in C} s + \max_{s \in C} s \le B,   \label{eq:CprimeValid}
			\end{align}
			as desired (the last inequality following from the validity of $C$).

			\item Otherwise $k \geq 2$. In this case, $C'$ contains $k-1 \geq 1$ blocks and the total size of $\tilde{S}$ demands in $C'$ is $(k-1) \cdot \e$, while the total size of $S$ demands in $C$ is at least $k \cdot \e$. Since both configurations have the same $\tilde{M}$ demands, this gives $\sum_{s \in C'} s \leq \sum_{s \in C}  s - \e$. Further, because $\tilde{S}$ demand has size exactly $\e$, we have $$\max_{s \in C'} s \le \max\bigg\{\e\,,\, \max_{s \in C' \cap \tilde{M}} s\bigg\} = \max\bigg\{\e\,,\, \max_{s \in C \cap \tilde{M}} s\bigg\} \le \e + \max_{s \in C} s.$$ Combining both bounds gives the desired inequality (as in \eqref{eq:CprimeValid}).
		\end{enumerate}
		Thus $C'$ is a valid configuration, concluding the proof of the proposition. 
	\end{proof}

	Finally, we construct a feasible setting of the $x$-variables for $(\lpconfigmin(\tilde{S} \cup \tilde{M}))$ with objective value at most $\big(1 + O(\epsilon)\big) \lpconfigmin(S \cup \tilde{M}) + O(1)$, completing the proof. Let $x^*$ be an optimal solution for $(\lpconfigmin(S \cup \tilde{M}))$. Let $\C$ be the collection of all possible configurations for demands $S \cup \tilde{M}$ and $\C'$ the collection of all possible configurations for demands $\tilde{S} \cup \tilde{M}$. For every $C' \in \C'$, we define $\bar{x}_{C'} = \sum_{C \in \C \mid C \rightarrow C'} x^*_C$, where $C \rightarrow C'$ denotes the event that the above procedure maps configuration $C$ to $C'$. In words, we map the configurations chosen by $x^*$ in $C$ to configurations in $C'$. At this point, the objective value of $\bar{x}$ is $\sum_{C' \in \C'} \bar{x}_{C'} = \lpconfigmin(S \cup \tilde{M})$. It remains to modify $\bar{x}$ so that it is feasible.
	
	Note that $\lpconfigmin(\tilde{S} \cup \tilde{M})$ and $\lpconfigmin(S \cup \tilde{M})$ have the same constraints for the demands in $\tilde{M}$, and the mapping $C \rightarrow C'$ preserves the number of each type of medium demand. Thus, $\bar{x}$ satisfies all medium demand constraints. It remains to satisfy the constraint for the demands in $\tilde{S}$.
	
	This constraint says that the total number of demands from $\tilde{S}$ in the configurations picked by the solution has to be at least twice the number of demands in $\tilde{S}$. Since every demand in $\tilde{S}$ has exactly the same size $\e$, this constraint can be equivalently written in terms of sizes as follows (let $\size(A) := \sum_{s \in A} s$ for $A \subseteq \R_+$): 
	\begin{align}
	\frac{1}{\e} \sum_{C' \in \C'} x_{C'} \cdot \size(C' \cap \tilde{S}) \,\ge\, 2 |\tilde{S}|. \label{eq:equivConstr}
	\end{align}
	We modify $\bar{x}$ to satisfy this. 
		
	We know that since $x^*$ is feasible for $(\lpconfigmin(S \cup \tilde{M}))$, the total size of $S$ demands it picks up is at least twice $\size(S)$, that is, 
	\begin{align}
	\sum_{C \in \C} x^*_C \cdot \size(C \cap S)  \,\ge\, 2 \size(S). \label{eq:equivConstr2}
	\end{align}
	Also, notice that for every mapped configurations $C \rightarrow C'$ we have $\size(C \cap S) \le \size(C' \cap \tilde{S}) + 2 \e$ (since $\size(C \cap S) \le (k+1) \cdot \e$ and $\size(C' \cap \tilde{S}) \ge (k-1)\cdot \e$ for some $k$). Then using the definition of $\bar{x}$ we get 
	 \begin{align*}
	 	\sum_{C' \in \C'} \bar{x}_{C'} \cdot \size(C' \cap \tilde{S}) = \sum_{C' \in \C'} \sum_{C \in \C \mid C \rightarrow C'} x^*_{C} \cdot \size(C' \cap \tilde{S}) &\ge \sum_{C \in \C} x^*_C \cdot \size(C \cap S) - 2 \e \sum_{C \in \C} x^*_C\\
	 	&\ge 2\size(S) - 2 \e\, \lpconfigmin(S \cup \tilde{M}),
	 \end{align*}
	 where the last inequality uses \eqref{eq:equivConstr2}. Thus, it suffices to increase the $\tilde{S}$ size of the demands picked up by $\bar{x}$ by $2\e \cdot (1 + \lpconfigmin(S \cup \tilde{M}))$ for it to satisfy \eqref{eq:equivConstr}. Dividing through by $\e$ and using $|\tilde{S}| =  \lceil \frac{1}{\e} \sum_{s \in S} s \rceil \le \frac{\size(S)}{\e} + 1$ we get 
	 \begin{align*}
	 \frac{1}{\e} \sum_{C' \in \C'} \bar{x}_{C'} \cdot \size(C' \cap \tilde{S}) \,\ge\, 2 |\tilde{S}| - 2 -  2 \lpconfigmin(S \cup \tilde{M}).
	 \end{align*}
	For $\bar{x}$ to satisfy \eqref{eq:equivConstr} we just need to increase it so it covers $\lceil 2 +  2 \lpconfigmin(S \cup \tilde{M}) \rceil$ extra demands of (type) $\tilde{S}$. For that, use \Cref{lem_matching_unit} to assign these additional demands to at most $O(\e) \cdot (1 + \lpconfigmin(S \cup \tilde{M})) + O(1)$ many machines, and add to $\bar{x}$ the configurations of these machines. Now $\bar{x}$ satisfies \eqref{eq:equivConstr}, and has objective value at most $\big(1 + O(\e)\big) \lpconfigmin(S \cup \tilde{M}) + O(1)$.
\end{proof}


\subsection{Putting it all together}

	We finally obtain the complete algorithm for assigning the original demands $J$. At a high-level, we schedule the large demands $L$ in edges by themselves, use the configuration LP plus rounding and realization of the configurations into an edge assignment (\Cref{thm_lpconfig_main}) for the instance $\tilde{J}$ with modified small and medium demands to create a template, and finally replace them by the original small and (non-large) medium items $S$ and $M \setminus L$. The precise algorithm is the following:

\begin{algorithm}
\small
\caption*{\textbf{\minoffalg:} Given a collection $J$ of demands, failover capacity $B \geq 1$, and parameter $\e \in (0,1)$:}\label{alg:minoffalg}
\begin{algorithmic}[1]
	\itemsep=1.5ex
	
	\vspace{1ex}
	
	\State Construct the sets of small, medium, and large demands $J = S \cup M \cup L$ as well as the blocks of small demands and grouped medium demands $\tilde{S}$ and $\tilde{M}$, respectively as in \Cref{sec_small_large_jobs}.

	\State For each large demand, open a new edge (two machines) for that demand and assign it there.

	\State For the blocks and grouped medium demands $\tilde{S} \cup \tilde{M}$, there is a total of $T = O(\frac{1}{\e^3})$ demand types and each configuration can have at most $D = O(\frac{1}{\e^2})$ demands. So run the algorithm guaranteed by \Cref{thm_lpconfig_main} to obtain an assignment of $\tilde{S} \cup \tilde{M}$ into at most $\lpconfigmin(\tilde{S} \cup \tilde{M}) + O(\frac{1}{\e^5})$ machines. We open this many machines and use the above assignment as a template to actually assign $S$ and $M \setminus L$

	\State Recall that we only increased the size of each demand from $M \setminus L$ to $\tilde{M}$, so we can assign each demand in $M \setminus L$ in the place of its corresponding demand in $\tilde{M}$.

	\State For the demands in $S$, we consider them in arbitrary order. When considering a demand $s \in S$, if there exists a block in the template (i.e. relative to a demand from $\tilde{S}$) assignment with less than $\e - s$ units of small demands assigned there, then assign demand $s$ in this block. If no such block exists, then we assign all remaining small demands using \Cref{lem_matching_unit}.
\end{algorithmic}
\end{algorithm}

\begin{proof}[Proof of \Cref{thm_main_off_min}]
	It is immediate that \minoffalg runs in polynomial time and assigns all demands. Further, this assignment satisfies the Nominal and Failover constraints: it puts each large demand on its own matching edge, our template assignment of $\tilde{S} \cup \tilde{M}$ is feasible by \Cref{thm_lpconfig_main}, we only assign smaller demands in $S \cup M$ than in the template, and our assignment of the remaining small demands is feasible by \Cref{lem_matching_unit}.
	
	It remains to show that the number of machines used is at most $\big(1 + O(\epsilon)\big) \lpconfigmin(J) + O(\frac{1}{\epsilon^5})$. We account the machines for the large, medium, and small demands separately.
	
	\begin{enumerate}
		\item For the large demands, we open $O(\lvert L \rvert) = O(\e^3 |M|)$ machines. Moreover, every demand in $M$ has size at least $\e^2$, so every feasible configuration for the demands $J$ has at most $\frac{1}{\e^2}$ such medium demands; since $\lpconfigmin(J)$ needs to pick enough configurations to cover twice the medium demands, we get that $2|M| \le \lpconfigmin(J) \cdot \frac{1}{\e^2}$. Thus, $O(\e) \cdot \lpconfigmin(J)$ machines are opened for the large demands.

		\item For the medium demands, our template assignment of $\tilde{S} \cup \tilde{M}$ opens $\lpconfigmin(\tilde{S} \cup \tilde{M}) + O(\frac{1}{\e^5}) \leq \big(1 + O(\e)\big) \lpconfigmin(J) + O(\frac{1}{\e^5})$, using \Cref{lem_small_med_lp}. We assign all medium demands in these machines.
		
		\item For the small demands, it suffices to bound the number of extra machines needed for the remaining small demands that do not fit in the blocks. Consider the first time that we consider a small demand that cannot be assigned to any block. It must be the case that each block is already assigned at least $(\e - \e^2)$ units of small demands. Recall that the number of blocks is $\lceil \frac{1}{\epsilon} \sum_{j \in S} s_j \rceil$. Thus the total size of already assigned small demands is at least $(\e - \e^2) \lceil \frac{1}{\e} \sum_{s \in S} s \rceil \geq (1 - \e) \sum_{j \in S} s_j$. We conclude that the total size of remaining unassigned small demands is at most $\e \cdot \sum_{s \in S} s = O(\e) \cdot \lpconfigmin(J)$ (again $\lpconfigmin(J)$ needs to cover twice all demands in $S$ and each unit of configuration picked by $J$ can cover at most 1 unit of size of these demands); this requires $O(\e) \cdot \lpconfigmin(J)$ machines by \Cref{lem_matching_unit}.
	\end{enumerate}
	
	In total, we see that the algorithm uses at most $\big(1 + O(\epsilon)\big) \lpconfigmin(J) + O(\frac{1}{\epsilon^5})$ machines as claimed. This concludes the proof of Theorem \ref{thm_main_off_min}.
\end{proof}

\ifx \hasmain \undefined
	\end{document}
\fi

\section{Rate of Convergence of the Minimum Number of Machines}
\label{sec:asConverge}

    \mnote{Need to take a pass on the whole section, adding text and making notation consistent}
    
    The goal of this section is to understand the minimum number of machines needed to assign $T$ items drawn i.i.d. from distribution $\mu$. Recall we denote this random variable by $\mach(X_1, \ldots, X_T)$. Our main result here (\Cref{thm:converges}, restated) is that in expectation, this random variable is approximately linear in $T$.
    
    \thmconverges*

	One should interpret the constant $c(\mu)$ as the average number of machines needed per demand as the number of demands goes to infinity. Thus, by dividing both sides of the theorem by $T$, we have a quantitative convergence for the expected average number of machines needed per demand for $T$ demands, $\frac{1}{T}\, \E\, \mach(X_1,\ldots,X_T) \,$, to the limiting value $c(\mu)$.
	
	The main idea to prove \Cref{thm:converges} is to consider a deterministic proxy for $\mach(X_1, \dots, X_T)$. To construct this proxy we follow the approach of~\cite{binPackingKnownT} that proves a similar result for the Bin Packing problem: for any distribution $\mu$ supported on $[0,1]$, we define its \emph{quantile function} $\mu^{-1}$ $$\mu^{-1}(p) := \inf\{ x \in [0,1] : \mu([0,x]) \ge p\}.$$ For example, if $\mu$ is a continuous distribution, then $\mu^{-1}(p)$ is the unique value $x$ such that $\mu([0,x]) = p$. 
	Then, let $\mu_T$ denote the instance that has $T$ demands whose sizes are given by $$\bigg\{\imu\bigg(\frac{0}{T}\bigg), \imu\bigg(\frac{1}{T}\bigg), \ldots, \imu\bigg(\frac{T-1}{T}\bigg)\bigg\}.$$
	Note that $\mu_T$ is a deterministic instance. Roughly, in the stochastic instance $\{X_1, \ldots, X_T\}$, we ``expect'' one demand to fall into each quantile $[\mu^{-1}(\frac{k}{T}), \mu^{-1}(\frac{k+1}{T})]$ for each $k = 0, \dots, T-1$. Thus, our deterministic proxy for $\mach(X_1, \dots, X_T)$ is $\mach(\mu_T)$.
	
	Keeping in mind our interpretation for the constant $c(\mu)$ from before (the average number of machines needed per demand) and our deterministic proxy, we take $c(\mu) := \limsup_{n \rightarrow \infty} \frac{1}{n} \OPT(\mu_n)$. We will show that this choice of $c(\mu)$ has the desired property.

	There are two main steps to prove \Cref{thm:converges}. We first show that $\mach(\mu_T)$ is a good proxy for $\E \, \mach(X_1, \ldots, X_T)$. The proof of the next lemma relies on another Rhee-Talgrand-like monotone matching argument, where we construct a matching between the $X_t$'s and $\mu^{-1}(\frac{t}{T})$'s such that few demands are left unmatched.
	
	
	\begin{lemma}\label{lemma:optQuantile}
		For every $T \in \mathbb{N}$, we have $$\E \, \mach(X_1,\ldots,X_T) \in \mach(\mu_T) \pm O(\sqrt{T}),$$ where $X_1,\ldots,X_T$ are i.i.d. samples from $\mu$. 
	\end{lemma}

	Second, we show that $\mach(\mu_T)$ has the desired approximate linearity property. This relies on relating $\mach(\mu_T)$ to its LP relaxation, $(\lpconfigmin(\mu_T))$, whose optimal value is approximately linear in $T$.
	
	\begin{lemma}\label{lemma:quantile}
		For every $T \in \mathbb{N}$, we have $\mach(\mu_T) \in T \cdot c(\mu) \pm O(T^{5/6})$.
	\end{lemma}

	\Cref{thm:converges} follows immediately from the above two lemmas, which we prove in the subsequent sections.

\subsection{Proof of \Cref{lemma:optQuantile}: $\mach(\mu_T)$ is a good proxy}

	For convenience, we let $\mu_T = \{s_0, \dots s_{T-1}\}$, where $s_j = \mu^{-1}(j/T)$. There are two analogous directions to prove: $\E \, \mach(X_1,\ldots,X_T) \leq \mach(\mu_T) + O(\sqrt{T})$, and
	$\mach(\mu_T) \leq \E \, \mach(X_1,\ldots,X_T) + O(\sqrt{T})$. For the former, we use the assignment of $\mu_T$ into $\mach(\mu_T)$ devices as a template to assign $X_1, \dots, X_T$ using only $O(\sqrt{T})$ extra devices. To do so, we show that there exists a large \emph{monotone matching} from the $X_i$'s to the $s_j$'s such that if $X_i$ is matched to $s_j$, then $X_i \leq s_j$. We can bound the number of unmatched $X_i$'s using a quantitative version of Hall's theorem (Theorem 1.3.1 of~\cite{lovaszPlummer}):
	
	\begin{thm}\label{thm:hall}
		Let $G = (L \cup R, E)$ be a bipartite graph. For any subset $U \subseteq L$, we define its \emph{deficiency} by $def(U) := \lvert U \rvert - \lvert N(U) \rvert$, where $N(U) \subseteq R$ is the set of neighbors of $U$. Then there exists a matching in $G$ that leaves at most $\max_{U \subseteq L} def(U)$ vertices of $L$ unmatched.
	\end{thm}

	For all matched $X_i$'s, we can assign them to the same position as their matched $s_j$-counterpart using $\mach(\mu_T)$ machines as a template. For the unmatched ones, we assign them each to their own disjoint edge (opening $2$ extra devices). To complete the proof, we need to show that in expectation, few of the $X_i$'s are unmatched. To do so, we use the Dvoretzky-Kiefer-Wolfowitz Inequality (\Cref{thm:DKW} in \Cref{app:conc}) to quantify the deviation of the empirical quantiles of the $X_i$'s with the ``true'' quantiles $\mu^{-1}(j/T)$. We now proceed formally.
	
	\begin{prop}\label{prop:optQuantileLe}
		We have $\E \, \mach(X_1,\ldots,X_T) \leq \mach(\mu_T) + O(\sqrt{T})$.
	\end{prop}
	\begin{proof}
		Consider the (random) bipartite graph $G$ with $T$ vertices on each side such that the left side vertices correspond to the $X_i$'s and the right to the $s_j$'s. We have an edge $(X_i, s_j)$ exactly when $X_i \leq s_j$.
		
		Note that the maximum deficiency subset of the $X_i$'s (as defined in \Cref{thm:hall}) must correspond to the random subset of all $X_i$'s that are strictly larger than some $s_{j-1}$ for some $j = 1, \ldots, T$. The deficiency of the $j$th such subset is $def(j) := \#\{\text{$X_i$'s strictly larger than $s_{j-1}$}\} - (T-j)$. Thus, a maximum matching in $G$ leaves at most $\max_j def(j)$ of the $X_i$'s unmatched by \Cref{thm:hall}. Fix some such maximum matching. We use it to assign the $X_i$'s as follows:
		\begin{enumerate}
			\item Open $\mach(\mu_T)$ many devices. Consider the tentative assignment of $\mu_T$ to these devices. \vspace{-3pt}
			\item For each matched $X_i$, we assign it to the pair of devices that its matched $s_j$ is tentatively assigned to. \vspace{-3pt}
			\item For each unmatched $X_i$, we open two more devices and assign $X_i$ to the edge between them.
		\end{enumerate}
		This is a feasible assignment of all $X_i$'s (because we assign each matched $X_i$ to a slot for a larger $s_j$  and each $X_i$ fits on an edge by itself) using at most $\mach(\mu_T) + 2\cdot \max_j def(j)$ devices. It remains to show $\E \, \max_j def(j) = O(\sqrt{T})$. We re-write $def(j)$ using \Cref{claim:quantileMeas} from \Cref{app:auxiliary}\mnote{Add intuition/blurb for this claim} in terms of the tails of $\mu$:
		\begin{align*}
			def(j) &= \#\{\text{$X_i$'s strictly larger than $s_{j-1}$}\} - (T-j) \\
			&\leq \#\{\text{$X_i$'s strictly larger than $s_{j-1}$}\} - T\cdot\mu((s_{j-1}, 1]) + 1.
		\end{align*}
		Moreover, using the Dvoretzky-Kiefer-Wolfowitz Inequality (\Cref{thm:DKW}) we have for any $\lambda > \frac{1}{T}$ 
		\begin{align*}
			\Pr\bigg(\frac{1}{T} \max_j def(j) \ge \lambda\bigg) &\le \Pr\bigg(\max_{v \in [0,1]} \bigg(\frac{1}{T}\,\textrm{\#\{$X_i$'s strictly bigger than $v$\}} - \mu((v,1]) \bigg) \ge \lambda -  \frac{1}{T} \bigg)\\
			&\le 2 e^{-2 T (\lambda - \frac{1}{T})^2}.
		\end{align*}
		Integrating the tail gives:
		\begin{align*}
			\E \, \frac{1}{T} \max_j def_j &\,\le\, O\bigg(\frac{1}{\sqrt{T}}\bigg) + \int_{2/\sqrt{T}}^\infty \Pr\bigg(\frac{1}{T} \max_j def(j) \ge \lambda\bigg) d\lambda\\
			&\,\le\, O\bigg(\frac{1}{\sqrt{T}}\bigg) + \int_{2/\sqrt{T}}^\infty 2e^{-2T(\lambda - \frac{1}{T})^2} d\lambda ~=~ O\bigg(\frac{1}{\sqrt{T}}\bigg),
		\end{align*}
		where the last inequality can be seen, for example, by noticing that the integral is at most a constant times the mean of a folded normal distribution with standard deviation $1/\sqrt{2T}$, which is $O(1/\sqrt{T})$. Re-arranging gives $\E \max_j (def(j)) \le O(\sqrt{T})$,\mnote{Check this plus!}\knote{Took the + out} as required.
	\end{proof}
	
	The proof of the other direction is analogous.
	
	\begin{prop}\label{prop:optQuantileGe}
	We have $\mach(\mu_T) \leq \E \, \mach(X_1,\ldots,X_T) + O(\sqrt{T})$.
	\end{prop}
	\begin{proof}
		We again consider a random bipartite graph $G$ on the same vertices but we switch the roles of the $X_i$'s and $s_j$'s. That is, now we have an edge $(s_j, X_i)$ exactly when $s_j \leq X_i$. Analogously, the maximum deficiency subset of the $s_j$'s corresponds to some set $\{s_j, \dots, s_{T-1}\}$ for some $j = 0, \ldots, T-1$ with deficiency $def(j) := T-j - \#\{\text{$X_i$'s at least $s_j$}\}$.
		
		As before, we use the tentative assignment of demands $X_1, \ldots, X_T$ into $\mach(X_1, \ldots, X_T)$ devices and a maximum matching of $G$ that leaves at most $\max_j def(j)$ of the $s_j$'s unmatched to assign $\mu_T$. In particular, we open $\mach(X_1, \ldots, X_T)$ devices and assign each matched $s_j$ to the slot of its matched $X_i$. For all remaining unmatched $s_j$'s, we assign them to disjoint edges. This gives a feasible assignment of the $s_j$'s into at most $\mach(X_1, \ldots, X_T) + 2 \cdot \max_j def(j)$ devices. It remains to show $\E \, \max_j def(j) = O(\sqrt{T})$.
		
		Again using \Cref{claim:quantileMeas} we have
		\[def(j) \le T \cdot \mu([s_j,1]) - \textrm{\#\{$X_i$'s at least $s_j$\}},\]
		and so again using the DKW Inequality we get $\Pr(\frac{1}{T} \max_j def(j) \ge \lambda) \le 2e^{-2T \lambda^2}$ for any $\lambda > 0$. An analogous calculation by integrating the tail gives $\E \, \max_j def(j) = O(\sqrt{T})$, as required.
	\end{proof}

	To summarize, in both directions (from the $X_i$'s to $s_j$'s and the reverse), we can use a monotone matching and template assignment to find a good assignment of one type of demands from the other. Combining both propositions proves \Cref{lemma:optQuantile}.

\subsection{Proof of \Cref{lemma:quantile}: Approximate linearity of $\mach(\mu_T)$}

Now we relate the optimum of the finite deterministic instances $\mu_T$ and the limit optimum $c(\mu) = \limsup_{n \rightarrow \infty} \frac{1}{n} \OPT(\mu_n)$. Again we have two directions to prove. For the first (more difficult) direction, we relate $\mach(\mu_T)$ with its LP relaxation, $\opt(\lpconfigmin(\mu_T))$.

\begin{prop} \label{prop:UBQuantile}
	For all $T \in \mathbb{N}$\replace{}{ and $\e \in (0,1)$}\knote{redundant $\epsilon$?} we have $$\mach(\mu_T) \leq T \cdot c(\mu) + O(T^{5/6}).$$
\end{prop}

\begin{proof}
	It suffices to prove for $\e \in (0,1)$:
	\begin{align}
		\mach(\mu_T) \le (1+O(\e))\,T \cdot \frac{\mach(\mu_{kT})}{kT} + O\bigg(\frac{1}{\e^5}\bigg) \label{eq:UBQuantile}
	\end{align}
	for all integers $k \ge 1$. Then taking $\limsup_{k\rightarrow \infty}$ on both sides and noticing $$\limsup_{k \rightarrow \infty} \frac{1}{kT} \mach(\mu_{kT}) \le \limsup_{n\rightarrow \infty} \frac{1}{n} \mach(\mu_n) = c(\mu)$$ gives $\mach(\mu_T) \leq (1+O(\e))\,T \cdot c(\mu) + O(\frac{1}{\e^5})$ (recall that passing to a subsequence cannot increase a $\limsup$). Setting $\e = \frac{1}{T^{1/6}}$ to optimize the bound gives
	\begin{align*}
		\mach(\mu_T) \,\le\, T \cdot c(\mu) + O(T^{5/6})
	\end{align*}		
	as desired. 
	%
	
	It remains to prove \eqref{eq:UBQuantile}. Recall that $\opt(\lpconfigmin(J))$ denotes the optimal value of the configuration LP \eqref{eq_lpconfigmin} for the set of demands $J$. Then \Cref{thm_main_off_min} allows us to bound the gap between $\mach(\mu_T)$ and $\opt(\lpconfigmin(\mu_T))$.
	\begin{align}
		\mach(\mu_T) \le (1+ O(\e))\, \opt(\lpconfigmin(\mu_T)) + O\bigg(\frac{1}{\e^5}\bigg).   \label{eq:UBQuantile2}
	\end{align}		
	Moreover, notice that $\opt(\lpconfigmin)$ is linear with respect to duplicating items. That is, for every integer $k$ we have
	\begin{align}
		\opt(\lpconfigmin(k \cdot \mu_T)) = k \cdot \opt(\lpconfigmin(\mu_T)),  \label{eq:UBQuantile3}
	\end{align}
	where $k \cdot \mu_T$ denotes the instance that has $k$ copies of each item in $\mu_T$.
	
	It remains to relate $k \cdot \mu_T$ with $\mu_{kT}$. Because the inverse CDF function $\imu$ is non-decreasing, we can relate these two sets of demand as follows.
	\begin{align*}
		\arraycolsep=1pt
		\begin{array}{rcccc}
			k\cdot \mu_T = \bigg\{&\underbrace{\imu\big(\tfrac{0}{T}\big),\ldots, \imu\big(\tfrac{0}{T}\big)}_{\textrm{$k$ times}}, &\underbrace{\imu\big(\tfrac{1}{T}\big),\ldots, \imu\big(\tfrac{1}{T}\big)}_{\textrm{$k$ times}},   &\ldots, &\underbrace{\imu\big(\tfrac{T-1}{T}\big),\ldots, \imu\big(\tfrac{T-1}{T}\big)}_{\textrm{$k$ times}}  \bigg\} \\
			\mu_{kT} = \bigg\{&\underbrace{\imu\big(\tfrac{0}{kT}\big),\ldots, \imu\big(\tfrac{k-1}{kT}\big)}_{\textrm{first $k$ items}}, &\underbrace{\imu\big(\tfrac{k}{kT}\big),\ldots, \imu\big(\tfrac{2k-1}{kT}\big)}_{\textrm{next $k$ items}},    &\ldots, &\underbrace{\imu\big(\tfrac{(T-1)k}{kT}\big),\ldots, \imu\big(\tfrac{Tk-1}{kT}\big)}_{\textrm{next $k$ items}}  \bigg\}			
		\end{array}
	\end{align*}
	we see that the sizes in $\mu_{kT}$ dominate those in $k \cdot \mu_T$ (i.e., there is a perfect monotone matching from $k\cdot \mu_T$ to $\mu_{kT}$). Then one can see that the optimal LP values for these instances satisfy the expected relationship $\opt(\lpconfigmin(k \cdot \mu_T)) \le 	\opt(\lpconfigmin(\mu_{kT}))$.
	Together with \eqref{eq:UBQuantile3} this gives $$\opt(\lpconfigmin(\mu_T)) = \frac{1}{k} \opt(\lpconfigmin(k \cdot \mu_T)) \leq \frac{1}{k} \opt(\lpconfigmin(\mu_{kT})).$$
	Combining this bound with \eqref{eq:UBQuantile2} completes the proof.
\end{proof}

Finally, we need a converse to the above proposition

\begin{prop} \label{prop:LBQuantile}
	For every $T \in \mathbb{N}$ we have 
	\begin{align*}
		\mach(\mu_T) \,\ge\, T \cdot c(\mu) - 2.
	\end{align*}
\end{prop}

\begin{proof}
	It suffices to show that for every $n$ we have 
	\begin{align}
		\mach(\mu_T) \ge T \cdot \frac{\mach(\mu_n)}{n} - 2 - \frac{2T^2}{n}. \label{eq:LBQuantile} 
	\end{align}
	Taking the $\limsup_{n \rightarrow \infty}$ gives the desired result.
	
	To prove \eqref{eq:LBQuantile}, we fix $n$ and write it as $n = kT + r$ for non-negative integers $k,r$ with remainder $r < k$. We will upper bound $\mach(\mu_n)$ as a function of $\mach(\mu_T)$. To do so, we first construct the intermediate instance $S$ obtained by increasing the size of the demands in $\mu_n$ as follows (recall that $\mu_n$ has items $\imu(\tfrac{j}{kT+r})$  for $j=0,\ldots ,kT+r-1$):
	
	\begin{itemize}
		\item For every $j = 0,\ldots,Tk-1$, let $\tilde{i}$ be such that $j \in [(\tilde{i}-1) k, \tilde{i} k)$; then take the item $\imu(\tfrac{j}{kT+r})$ of $\mu_n$ and increase its size to $\imu(\tfrac{\tilde{i}k}{kT})$, and add the latter to $S$.
		\item For every $j = Tk,\ldots,Tk + r - 1$, take the item $\imu(\tfrac{j}{kT+r})$ of $\mu_n$ and increase its size to $\imu(1)$, and add the latter to $S$. 
	\end{itemize}
	We have only increased demand sizes from $\mu_n$ to $S$, so we have $\mach(\mu_n) \leq \mach(S)$. To further upper bound $\mach(S)$, notice that $S$ has the structure 
	\begin{align*}
		S = \bigg\{ \underbrace{\imu\big(\tfrac{k}{kT}\big), \ldots, \imu\big(\tfrac{k}{kT}\big)}_{\textrm{$k$ times}}, \,\underbrace{\imu\big(\tfrac{2k}{kT}\big), \ldots, \imu\big(\tfrac{2k}{kT}\big)}_{\textrm{$k$ times}},\ldots,\,\underbrace{\imu\big(\tfrac{Tk}{kT}\big), \ldots, \imu\big(\tfrac{Tk}{kT}\big)}_{\textrm{$k$ times}},\,\underbrace{\imu(1), \ldots, \imu(1)}_{\textrm{$r$ times}}    \bigg\}, 
	\end{align*}
	which is exactly the union of $k$ copies of the instance $\overline{\mu_T} := (\mu_T \cup \{\imu(1)\} )\setminus \{\imu(0)\}$ and the instance of ``big'' demands $B$ that has $r$ items of size $\imu(1)$, i.e. $S = (k\cdot \overline{\mu_T}) \cup B$. Next, observe the subadditivity relation $$\mach(S) \le k \cdot \mach(\overline{\mu_T}) + \mach(B),$$ since the optimal solutions of each of the instances $\overline{\mu_T}$ and $B$ can be concatenated, giving a feasible solution for $S$ with $k \cdot \mach(\overline{\mu_T}) + \mach(B)$ machines. Moreover, we claim that 
	\begin{gather*}
		\mach(\overline{\mu_T}) \le \mach(\mu_T) + 2~~~~~\textrm{and}~~~~~\mach(B) \le 2r:
	\end{gather*}
	The first inequality is because we can assign all demands $\mu_T \setminus \{\imu(0)\}$ using at most $\mach(\mu_T)$ machines and then assign the remaining demand $\mu^{-1}(1)$ using $2$ extra machines; the second inequality is because a feasible solution for $B$ is to assign each demand to $2$ separate machines. 
	
	Putting all of these bounds together we obtain 
	\begin{align*}
		\mach(\mu_n) \le \mach(S) \le k\, \mach(\mu_T) + 2k + 2r.
	\end{align*} 
	Dividing though by $n$ and using the facts $\frac{k}{n} \le \frac{1}{T}$ and $\frac{r}{n} \le \frac{T}{n}$ we get
	\begin{align*}
		\frac{1}{n} \mach(\mu_n) \le \frac{1}{T} \mach(\mu_T) + \frac{2}{T} + \frac{2T}{n},
	\end{align*}
	which is equivalent to the desired inequality \eqref{eq:LBQuantile}. This concludes the proof. 
\end{proof}

Combining the above two propositions completes the proof of \Cref{lemma:quantile}. To summarize, for both propositions we needed the approximate linearity of $\mach(\mu_n)$. In the former, we argued via LP relaxations, and in the latter by concatenating sub-instances.

\bibliographystyle{acm}

\bibliography{bibliography}


\newpage
\appendix

\noindent {\LARGE \bf Appendix}

\section{Concentration Inequalities} \label{app:conc}
	
	We also need a couple of concentration inequalities, starting with McDiarmid's Inequality (Theorem 6.2 of \cite{lugosiConcentration}).
	
	\begin{lemma}[McDiarmid's Inequality] \label{lemma:mcDiarmid}
		Let $g : \mathcal{Z}^n \rightarrow \R$ be a function with the bounded differences property, i.e. for every two vectors $z, z' \in \mathcal{Z}^n$ that only differ in 1 coordinate we have $|g(z) - g(z')| \le M$. If $Z_1,\ldots,Z_n$ are independent random variables taking values in $\mathcal{Z}$, then for all $\alpha > 0$ $$\Pr(|g(Z) - \E g(Z)| \ge \alpha) \le  2 e^{-\frac{2 \alpha^2}{n M^2}},$$
	\end{lemma}
	
	The next classical inequality can be found for example in Theorem 2.8 of \cite{lugosiConcentration}.
	
	\begin{lemma}[Chernoff's Inequality] \label{lemma:chernoff}
		Let $Z_1,\ldots,Z_n$ be independent random variables in $[0,1]$. Then for all $\lambda > 0$
		\begin{align*}
			\Pr\bigg(\sum_{t \le n} Z_t - \E \sum_{t \le n} Z_t  \,\ge\, \lambda \bigg) \,\le\, e^{-\frac{2 \lambda^2}{n}},
		\end{align*}
		and the same holds for the lower tail, i.e. replacing ``\,$\ge \lambda$'' for ``\,$\le -\lambda$''.
	\end{lemma}
	
	We also need the Dvoretzky-Kiefer-Wolfowitz inequality that bounds the rate of uniform convergence of the empirical cdf to the true cdf. To state it, given a distribution $\mu$ over the reals, let $\cdf(x) = \Pr_{X \sim \mu}( X \le x)$ denote its cdf, and given i.i.d. samples $X_1,\ldots,X_n \sim \mu$ let $\cdf_n(x) = \frac{1}{n}\sum_{i \le n} \ones(X_i \le x)$ be the empirical cdf. The following version of the DKW inequality is Corollary 1 of~\cite{DKWMassart}.
	
	\begin{thm}[DKW Inequality] \label{thm:DKW}
	   For any distribution $\mu$, any number of samples $n$, and all $\lambda > 0$, $$\Pr\bigg(\max_{x \in \R} |\cdf_n(x) - \cdf(x)| \ge \lambda \bigg) \le 2 e^{-2 n \lambda^2}.$$
	\end{thm}



	\section{Auxiliary Results} \label{app:auxiliary}
	
	\begin{claim} \label{claim:quantileMeas}
		Consider any probability measure $\mu$ over $[0,1]$. Letting $s_j = \imu(j/T)$, we have for $j = 0,\ldots,T-1$:
		\begin{itemize}
			\item $T- j \ge T \cdot\mu((s_{j-1}, 1]) - 1$ \vspace{-3pt}
			\item $T -j \le T \cdot \mu([s_j, 1])$.
		\end{itemize}
	\end{claim}
	
	\begin{proof}
		For the first item, by definition of $s_{j-1}$ we have (using continuity of measures w.r.t. decreasing sets) $\mu([0,s_{j-1}]) \ge \frac{j-1}{T}$, which reorganizing gives $j \le T \cdot \mu([0,s_{j-1}]) + 1$; this implies $T - j \ge T(1 - \mu([0,s_{j-1}]) - 1$, which is the desired bound. 
		
		For the second item, we have (using continuity of measures for increasing  sets) $\mu([0,s_j)) \le \frac{j}{T}$, which gives $T - j \le T (1 - \mu([0,s_j)))$, which is exactly what we need. 
	\end{proof}

	
	\begin{lemma} \label{lemma:inversion}
	   Consider  non-negative constants $c,d$ and $\alpha \in (0,1)$. If  $m,t > 0$ satisfy $m \ge c t - (d t)^{\alpha}$, then
	   \begin{align}
	   t \le \max\bigg\{\frac{1}{c} \bigg(m + \bigg(\frac{m (d+1)}{c}\bigg)^\alpha\bigg)~,~\bigg(\frac{(d+1) d^{\alpha}}{c}\bigg)^{\frac{1}{1-\alpha}} \bigg\}. \label{eq:inversion}
	   \end{align}
	\end{lemma}
	
	\begin{proof}
	    Assume $t^{1-\alpha} > \frac{(d+1) d^{\alpha}}{c}$, otherwise we are done by the second term in the $\max$. Multiplying through by $t^{\alpha}$ we obtain $t > \frac{(d+1) (dt)^{\alpha}}{c}$. Rewriting the left-hand-side as $t= \big(\frac{(d+1)c}{c} - d\big) \cdot t$ and reorganizing gives $$\frac{d+1}{c} \Big(c t - (d t)^{\alpha} \Big) > d t.$$ Then the first term in the max in \eqref{eq:inversion} multiplied by $c$ is
	    \begin{align*}
	        m  + \bigg(\frac{m (d+1)}{c}\bigg)^\alpha \stackrel{\star}{\ge}  c t - (d t)^{\alpha} + \bigg(\frac{d+1}{c} \Big(c t - (d t)^{\alpha}\Big)  \bigg)^{\alpha} \ge ct - (dt)^\alpha + (dt)^\alpha = ct,
	    \end{align*}
	    where the inequality $\star$ follows from the assumption $m \ge ct - (dt)^{\alpha}$. Dividing through by $c$ proves \eqref{eq:inversion} (using the first term in the $\max$). 
	\end{proof}


  \section{Additional Results for Online Worst-Case} \label{app:onWCS}

	\subsection{Upper bound} \label{sec:onWCLB}
	
	In this section, we show an upper bound of $\frac{1}{2}$ for deterministic online algorithms for the \problem problem in the worst-case, proving that the algorithm we design in  \Cref{sec:onWCSNR} is essentially tight. 
	
	\begin{thm} \label{thm:upperBoundDet}
	No deterministic online algorithm can obtain competitive ratio better than $\frac{1}{2}$ for the \problem problem in the worst-case model, even when $B = \infty$.
	\end{thm}

	\begin{proof}

Consider an instance with 4 devices and let $\e > 0$ be a parameter. Then, the demands arrive as follows:
\rnote{not clear what is meant by parallel edges so I rewrote}
\vspace{-5pt}
\begin{itemize}
\item The first 2 demands have size $\e$.\vspace{-5pt}
\item If the first 2 demands are placed on the same edge, then there are 2 more demands of size $1-\e$.\vspace{-5pt}
\item If the first 2 demands are not placed on the same edge, then there is one more demand of size 1. 
\end{itemize}

	\paragraph{Case 1: the first 2 demands go on the same edge.}	In this case, we can only fulfill one of the demands of size $1-\e$ that can be placed on the two devices with no existing load, giving the algorithm total value of at most $1+\e$. \OPT can place all 4 demands, e.g., by placing the $\e$ and $1- \e$ demands alternatively on edges of a fixed $4$-cycle. This gives optimal value 2. So, the competitive ratio of the algorithm is $\frac{1+\e}{2}$.

	\paragraph{Case 2: either the first 2 demands go on disjoint edges or on edges that share an endpoint.} In this scenario, the algorithm cannot place the demand of size 1, while \OPT can place it by putting the two demands of size $\e$ on a single edge. The competitive ratio in this case is $\frac{2\e}{1+2\e}$.

 \paragraph{} Taking $\e \rightarrow 0$ gives the desired result. 

\end{proof}

\subsection{Algorithm for small demands} \label{sec:onWCSsmall}
    
    	We assume there is $L \in \mathbb{N}$ such that all demands have size $\leq \frac{1}{L}$. We will show that there exists an online algorithm with competitive ratio that goes to 1 as the size of the largest demand goes to 0. In particular, the claim is the following:
	
	
	\begin{restatable}{thm}{thmSmall} \label{thm:small}
     If for some $L \in \N$  \replace{that is a square}{}\mnote{Double check if there is any losses due to $\sqrt{L}$ not being integral, etc. and state what we assume of $L$} all demands have size at most $\frac{1}{L}$, there is an  online algorithm for \problem in the worst-case model that has competitive ratio at least $\left(1 - \min \big\{ \frac{3}{\sqrt{L}}, \frac{m}{L} \big\} \right)$. 
\end{restatable}

	\subsubsection{Algorithm} The algorithm opens cliques of machines and  schedules the demands on their edges using first-fit. In particular, for each demand, the algorithm first considers all edges of the first clique in order before continuing on the edges of the second clique (if one exists) and so on. 
	
	Specifically, the algorithm is the following:

	\begin{enumerate}
	    \item If $m < 3\sqrt{L}$, open a single clique containing all $m$ machines. Otherwise, when a new clique is needed, open a clique of size $\sqrt{L}$; if only $\sqrt{L'} < \sqrt{L}$ vertices remain, then open a $\sqrt{L'}$-clique.
	    \item When a demand arrives, assign it to an edge of an opened clique using first-fit making sure that for each clique of size $m'$ the total load on each edge is at most 
	    $$\alpha_{m'} = \min \bigg\{ \frac{B}{m'}, \frac{1}{m'-1} \bigg\}$$
	    If needed, open a new clique if possible.
	    \item If the demand cannot be scheduled, then stop. 
	\end{enumerate}
	
	\subsubsection{Analysis} 
	Given the capacities $\alpha_{m'}$ on each edge of a $m'$-clique, it is easy to check that the algorithm creates a valid placement that satisfies both Nominal and Failover capacities.   

	Assume that the algorithm was not able to place everything (otherwise it is \OPT). Upon termination, there is a demand of size at most $\frac{1}{L}$ that could not be scheduled, therefore for each clique of size $m'$ that the algorithm opened, all edges have load at least $$\alpha_{m'} - \frac{1}{L}$$
	
	As a result, the total load on all edges of a $m'$-clique is at least 
	\begin{align}
        \frac{1}{2} \cdot m' \cdot (m'-1) \left(\alpha_{m'} - \frac{1}{L}\right) \label{eq:small_dem_1}
	\end{align}
	
	We also know that \OPT can achieve at most 
	\begin{align}
	    \frac{1}{2} \cdot \min \{m, (m-1) B \}.
	\end{align}
	
	We now consider the following cases:
	
	\paragraph{Case 1: $m < 3\sqrt{L}$.} In this case, the algorithm opens a single clique of size $m$ and schedules load at least 
	\begin{align}
        \frac{1}{2} \cdot m \cdot (m-1) \left(\alpha_{m} - \frac{1}{L}\right) \label{eq:small_dem_2}
	\end{align}

    Depending on the value of $B$ and $m$, we have the following cases:
    
    \begin{itemize}
        \item If $m < (m-1) B$, then \OPT $\leq \frac{m}{2}$, and $\alpha_m = \frac{1}{m-1}$. From \eqref{eq:small_dem_2}, we have:
	
	\begin{align*}
		Alg ~&\ge~ \frac{1}{2} \cdot m \cdot (m-1) \left(\frac{1}{m-1} - \frac{1}{L}\right) 
		= \frac{m}{2} \left(1 - \frac{m-1}{L}\right) 
		\geq \OPT \left(1 - \frac{m}{L}\right).
	\end{align*}
	
	\item 	If $m \geq (m-1) B$, then \OPT $\leq \frac{1}{2}(m-1) B$,\knote{The 1/2 was missing here} and $\alpha_m = \frac{B}{m}$. From \eqref{eq:small_dem_2}, we have:
	
	\begin{align*}
		Alg ~&\ge~ \frac{1}{2} \cdot m \cdot (m-1) \left(\frac{B}{m} - \frac{1}{L}\right) 
		= \frac{(m-1) B}{2} \left(1 - \frac{m}{BL}\right) 
		\geq \OPT \left(1 - \frac{m}{L}\right).
	\end{align*}
    \end{itemize}

    In both cases, $Alg \geq \OPT \left(1 - \frac{m}{L}\right) \geq \left(1 - \min \Big\{ \frac{3}{\sqrt{L}}, \frac{m}{L} \Big\} \right)$.

	\paragraph{Case 2: $m \geq 3\sqrt{L}$.} Let $C$ denote the number of $\sqrt{L}$-cliques that the algorithm opens. By construction, $C \cdot \sqrt{L} + \sqrt{L'} = m$.
	
	\paragraph{$\sqrt{L}$-cliques.} From \eqref{eq:small_dem_1}, the overall load that the algorithm scheduled successfully on these cliques is at least
	\begin{align}
		C\cdot \frac{1}{2} \cdot \sqrt{L} \cdot (\sqrt{L}-1) \left(\alpha_{\sqrt{L}} - \frac{1}{L}\right) \geq C \frac{\sqrt{L} \cdot (\sqrt{L}-1)}{2} \left(\frac{1}{\sqrt{L}} - \frac{1}{L}\right) \geq \frac{1}{2}  C  \sqrt{L}  \left(1 - \frac{2}{\sqrt{L}} \right). \label{eq:small1}
	\end{align}
	
    \paragraph{$\sqrt{L'}$-clique.} Similarly, for the $\sqrt{L'}$-clique, the total load that is scheduled on this clique is at least
	\begin{align}
			\frac{1}{2} \cdot \sqrt{L'} \cdot (\sqrt{L'}-1) \left(\alpha_{\sqrt{L'}} - \frac{1}{L}\right) \geq \frac{\sqrt{L'} \cdot (\sqrt{L'}-1)}{2} \left(\frac{1}{\sqrt{L'}} - \frac{1}{L}\right) \geq \frac{1}{2}    \sqrt{L'}  \left(1 - \frac{\sqrt{L'}}{L} - \frac{1}{\sqrt{L'}} \right). \label{eq:small2}
	\end{align}
	
	Combining the above, the algorithm gets value at least
	\begin{align*}
		Alg ~&\ge~ \frac{1}{2}  C  \sqrt{L}  \left(1 - \frac{2}{\sqrt{L}} \right) + \frac{1}{2}    \sqrt{L'}  \left(1 - \frac{\sqrt{L'}}{L} - \frac{1}{\sqrt{L'}} \right) \\
		&= \frac{1}{2} \left(m - 2\frac{m}{\sqrt{L}} + 2\frac{\sqrt{L'}}{\sqrt{L}}  - \frac{L'}{L} - 1  \right) \\
		&\geq \frac{1}{2} \left(m - 2\frac{m}{\sqrt{L}} - 1 \right) 
	\geq \frac{m}{2} \left(1 - \frac{3}{\sqrt{L}}  \right) \\
		&\geq \OPT  \left(1 - \frac{3}{\sqrt{L}}  \right) \geq \left(1 - \min \Big\{ \frac{3}{\sqrt{L}}, \frac{m}{L} \Big\} \right)
	\end{align*}
	where we used the fact that $C \cdot \sqrt{L} + \sqrt{L'} = m \Leftrightarrow C =  \frac{m-\sqrt{L'}}{\sqrt{L}} $, $L' < L $, and $\frac{m}{\sqrt{L}} \geq 1$.


\section{Omitted Proofs from Section \ref{sec:stoch}} \label{app:stoch_proofs}

\subsection{Proof of \Cref{lemma:stochDensityOPT}} \label{app:proofStochDensityOPT}

    In order to upper bound the utilization that \opt can achieve, the first step is to bound the minimum number of machines that are required to schedule a set of demands (with high probability). This allows us to generate bounds for the number of demands that can be scheduled (with high probability) when the number of machines is fixed. Connecting this number of demands with their sizes produces the desired bounds for \opt's utilization.

	In particular, recall that $\mach(s_1,\ldots,s_n)$ denotes the minimum number of devices needed to schedule all the demands $s_1,\ldots,s_n$ satisfying the Nominal and Failover constraints. Also recall that \Cref{thm:converges} shows that there exists a scalar $c(\mu)$ such that for every $T$, we have $$\E\, \mach(S_1,\ldots,S_T) \,\in\, T \cdot c(\mu) \pm O(T^{5/6}).$$

	However, for proving \Cref{lemma:stochDensityOPT}, we need the observation that this bound holds not only in expectation but with high probability (with a negligible additional loss). To see that, notice that the function $\mach(s_1,\ldots,s_n)$ has bounded differences: changing the size of any demand $i$ from $s_i$ to $s'_i$ (both in the range $[0 , \min\{1, \frac{B}{2}\}]$) can change the minimum number of machines required by at most 2, i.e., $$|\mach(s_1,\ldots,s_i,\ldots,s_n) - \mach(s_1,\ldots,s'_i,\ldots,s_n)| \le 2,$$ since we can always schedule the demand $i$ on an edge by itself (using 2 new machines) if needed. Then as a consequence of McDiarmid's Inequality (\Cref{lemma:mcDiarmid}) with $\alpha = \lambda \sqrt{T}$ we directly obtain the following.
	
	\begin{cor}	\label{cor:convergesWHP}
		For every $T$ we have $$\mach(S_1,\ldots,S_T) \,\in\, T \cdot c(\mu) \pm O(T^{5/6}) \pm \lambda \sqrt{T}$$ with probability at least $1 - 2e^{-\frac{\lambda^2}{2}}$.
	\end{cor}

	Then by essentially inverting this bound, we can upper bound how many demands are scheduled by the optimal solution $\OPT(S_1,\ldots,S_n)$ for our original problem. 

%
%
	
	\begin{lemma} \label{lemma:numScheduled}
		Consider the original problem \problem with $m$ machines. For every $\delta > 0$, with probability at least $1 - \delta$ the optimal solution to the instance $S_1,\ldots,S_n \sim \mu$ schedules at most $\frac{m}{c(\mu)} + O(m^{5/6}) + \log^{3/2} \frac{1}{\delta}$ demands. 
	\end{lemma}
	
	\begin{proof}[Proof sketch]
 Using \Cref{cor:convergesWHP} with $\lambda = \sqrt{2 \log \frac{2}{\delta}}$, one can see that there is a positive constant $d$ such that with probability at least $1-\delta$
	\begin{align}
	\mach(S_1,\ldots,S_T) \,\ge\, T \cdot c(\mu) - d^{5/6} \cdot T^{5/6} \label{eq:numScheduled}
	\end{align}
	for every number of demands $T \ge \log^{3/2} \frac{1}{\delta}$.
	
	A bit of algebra shows that there is a value $\bar{T} = \frac{m}{c(\mu)} + O(m^{5/6}) + \log^{3/2} \frac{1}{\delta}$ such that the right-hand side of \eqref{eq:numScheduled} is strictly more than $m$ (the term $+\log^{3/2} \frac{1}{\delta}$ being present just to ensure $\bar{T} \ge \log^{3/2} \frac{1}{\delta}$); in fact, taking $$\bar{T} := \max\bigg\{\frac{1}{c(\mu)}\bigg(m + \bigg(\frac{m (d+1)}{c(\mu)}\bigg)^{5/6}\bigg)\,,\, \frac{(d+1)\, d^{5/6}}{c(\mu)}\,,\, \log^{3/2} \frac{1}{\delta} \bigg\} + 1$$\knote{does the second term in the max need the exponent $1/(1-a)$ from the lemma?} suffices, which can be verified by applying the contrapositive of \Cref{lemma:inversion}). This means that with probability at least $1-\delta$, the demands $S_1,\ldots,S_{\bar{T}}$ cannot all be scheduled within $m$ machines; in such scenarios the optimal solution then schedules at most $\bar{T} - 1 = \frac{m}{c(\mu)} + O(m^{5/6}) + \log^{3/2} \frac{1}{\delta}$ demands, as claimed.  
	\end{proof}
	
	We can now show that with high probability $\OPT(S_1,\ldots,S_n) \le m \cdot \frac{\E S_0}{c(\mu)} + O(m^{5/6})$ and conclude the proof of \Cref{lemma:stochDensityOPT}. Setting $\delta = \frac{1}{m^2}$ and letting $\bar{T} = \frac{m}{c(\mu)} + O(m^{5/6})$ be the above upper bound on the number of demands scheduled by $\OPT(S_1,\ldots,S_n)$, with probability $\ge 1 - \frac{1}{m^2}$ we have that $\OPT(S_1,\ldots,S_n) \le \sum_{t \le \bar{T}} S_t$. Moreover, employing the Chernoff bound (\Cref{lemma:chernoff}) with $\lambda = \sqrt{\bar{T} \log m}$, with probability at least $1- \frac{1}{m^2}$ this sum can be upper bounded as 
	\begin{align*}
		\sum_{t \le \bar{T}} S_t \,\le\, \bar{T} \cdot \E S_0 + \sqrt{\bar{T} \log m} \,\le\, m \cdot \frac{\E S_0}{c(\mu)} + O(m^{5/6}). 
	\end{align*}
	Taking a union bound to combine the two previous bounds, with probability at least $1 - \frac{2}{m^2}$ we have that $\OPT(S_1,\ldots,S_n) \le  m \cdot \frac{\E S_0}{c(\mu)} + O(m^{5/6})$, as desired. This concludes the proof of the \Cref{lemma:stochDensityOPT}. 


	\subsection{Proof of \Cref{lemma:stochLogCalls}}	
	
	To prove \Cref{lemma:stochLogCalls} it will suffice to show that with high probability \oneround consumes a quarter of the machines available, unless there are already few machines available. 
		
	\begin{lemma} \label{lemma:openCst}
		There is a constant $\cstg$ such that whenever $\tilde{m} \ge 4 \cstg \cdot m^{5/6} + 8$, we have that $\oneround(\tilde{m})$ opens at least $\frac{\tilde{m}}{4}$ machines with probability at least $1 - \frac{\log(m/c(\mu))}{m^2}$.   
	\end{lemma}	

	\begin{proof}
	The main element of the proof is a converse to \Cref{claim:numPhases}, that is, a lower bound on the number of phases performed by \oneround. To argue that it has not ran out of machines on an initial phase, let $UB_\ell$ be the upper bound on the number of machines opened by the \oneround until phase $\ell$ given by \eqref{eq:openEll2}, namely $$UB_\ell = 2 n_\ell \cdot c(\mu) + \cstf \cdot n_{\ell}^{5/6} + \cstf \cdot m^{5/6}$$ for a sufficiently large constant $\cstf$. Here is the desired bound on the number of phases performed. 
	
		\begin{claim} \label{claim:LBPhases}
		Let $\underbar{k}$ be the largest integer $\ell$ such that $UB_\ell + 2 m^{5/6} \le \tilde{m}$. Then with probability at least $1 - \mubar{k}/m^2$ 
		the algorithm $\oneround(\tilde{m})$ performs at least $\mubar{k}$ phases. 
	\end{claim}
	
	\begin{proof}
	\mnote{This claim could be streamlined, feels clunky} 
	From \eqref{eq:openEll2} and the fact that $\cstf$ and $m$ are at least a sufficiently large constant, we have with probability at least $1-\mubar{k}/m^2$
	\begin{align}
	Open_{\mubar{k}} \,+\, \csta \cdot \sqrt{n_{\mubar{k}}}\, \log^{3/4} n_{\mubar{k}} \,\le\, UB_{\mubar{k}}\,. \label{eq:mubar}
	\end{align}
	Under this event we see that \oneround starts phase $\mubar{k}$, that is, it does not STOP in Step (b) in the beginning of this iteration, since at that point
	\begin{align*}
&~~\textrm{\#\,already open machines} + \overline{\mach}(Y_1,\ldots,Y_{n_k}) + \csta \cdot \sqrt{n_k}\, \log^{3/4} n_k + 2 m^{5/6}\\
&= Open_{\mubar{k}-1} \,+\, M_k \,+\, \csta \cdot \sqrt{n_k}\, \log^{3/4} n_k + 2 m^{5/6}	\\
&\le Open_{\mubar{k}} \,+\, \csta \cdot \sqrt{n_k}\, \log^{3/4} n_k \,+\, 2 m^{5/6}\\
&\le UB_{\mubar{k}} + 2m^{5/6} \le \tilde{m},
	\end{align*} 
	where the last inequality follows from the definition of $\mubar{k}$. Moreover, under this event the algorithm also does not run out of machines (i.e. fails) on this phase $\mubar{k}$, since again $Open_{\mubar{k}} \le UB_{\mubar{k}} \le \tilde{m}$. The claim then follows.
	\end{proof}

	Using this claim plus the lower bound on the number of machines $M_k$ open on a round $k$ from \Cref{claim:MU}, we have that with probability at least $1 - (\mubar{k}+1)/m^2$
	\begin{align}
		Open \ge Open_{\mubar{k}} \ge \sum_{k = k_0}^{\mubar{k}} M_k &\ge \sum_{k = k_0}^{\mubar{k}}  \bigg(n_k \cdot c(\mu) \,-\, 2  \csta \cdot n_k^{5/6}\bigg) \notag\\
		&\ge \sum_{k=1}^{\mubar{k}} n_k \cdot c(\mu) - \sum_{k=1}^{k_0} n_k \cdot c(\mu) - \sum_{k=1}^{\mubar{k}} 2 \csta \cdot n_k^{5/6}  \notag\\
		&\ge (2 n_{\mubar{k}} - 1)\,c(\mu) - 2 n_{k_0}  - 5 \csta \cdot n_{\mubar{k}}^{5/6} \notag \\
		&\ge (2 n_{\mubar{k}} - 1)\,c(\mu) - 2 m^{5/6}  - \frac{5 \csta}{c(\mu)^{5/6}} \cdot m^{5/6}, \label{eq:LBOpen}
	\end{align}
	\knote{redundant 2 in first line in front of cst1 but not that important...}
	where the next-to-last equality follows from the definition $n_k = 2^k$, and the last inequality uses \Cref{claim:MU} and that the definition of $\mubar{k}$ implies $n_{\mubar{k}} \le \frac{\tilde{m}}{c(\mu)} \le \frac{m}{c(\mu)}$.
	
 	To further lower bound this quantity, from the maximality of $\mubar{k}$ we have $UB_{\mubar{k}+1} + 2m^{5/6} > \tilde{m}$, which expanding the definition of $UB_{\mubar{k}+1}$ and again using $n_{\underbar{k}} \le \frac{m}{c(\mu)}$ gives
 	\begin{align*}
 		2 n_{\mubar{k}} \cdot c(\mu) > \frac{1}{2} \bigg(\tilde{m} - \cstf \cdot n^{5/6}_{\mubar{k}+1} - (\cstf + 2) m^{5/6}\bigg) \ge \frac{\tilde{m}}{2} -\bigg(\frac{\cstf}{2^{1/6} c(\mu)^{5/6}} + \frac{\cstf}{2} + 1\bigg)
 		m^{5/6}.
 	\end{align*} 
 	Employing this on inequality \eqref{eq:LBOpen} we get that with probability at least $1- 2\mubar{k}/m^2$ 
 	\begin{align*}
 	Open \ge \frac{\tilde{m}}{2} - \underbrace{\bigg(\frac{\cstf/(2^{1/6}) + 5 \csta }{c(\mu)^{5/6}} + \frac{\cstf}{2} + 3 \bigg)}_{=:\cstg}\, m^{5/6} - 2.
 	\end{align*}
	Under the assumption $\tilde{m} \ge 4 \cstg \cdot m^{5/6} + 8$ of \Cref{lemma:openCst}, we get $Open \ge \frac{\tilde{m}}{4}$ as desired. To finalize, again since $n_{\mubar{k}} \le \frac{m}{c(\mu)}$, we have $\mubar{k} \le \log \frac{m}{c(\mu)}$, and so this happens with probability at least $1 - \frac{\log (m/c(\mu))}{m^2}$\knote{needs a 2 in front or is fine?}. This concludes the proof of \Cref{lemma:openCst}. 
 \end{proof}
 
 We now turn to the proof of the main lemma. 
 
 \begin{proof}[Proof of \Cref{lemma:stochLogCalls}]
 	We assume that $m$ is at least a sufficiently large constant, otherwise the result directly holds \knote{*}(with an appropriate constant in the term $O(\frac{1}{m})$); in particular, we assume that $5 \cstg \cdot m^{5/6} \ge 4 \cstg \cdot m^{5/6} + 8$, which will be useful to clean up the bounds. 
 	
 	Let $\tilde{m}_i$ denote the number of unopened machines right before the $i$-th call to \oneround made by the main algorithm (i.e., at this call $\tilde{m}_i$ is the parameter passed to \oneround). Define the (bad) event $E_i$ that in the beginning of round $i$ we still have more than $5 \cstg \cdot m^{5/6}$ unopened machines but (unlike what is prescribed by \Cref{lemma:openCst} above) we did not consume open at least a quarter of these machines, i.e.  
 	\begin{align*}
 		E_i \equiv (\tilde{m}_i > 5 \cstg \cdot  m^{5/6}) \textrm{ and } (\tilde{m}_{i+1} > \tfrac{3}{4} \tilde{m}_i).
 	\end{align*}
 	
 	We claim that when neither of the events $E_1,\ldots,E_{\bar{r}}$ holds (for $\bar{r} := \frac{\log m}{\log 4/3}$), then the total number of machines opened by the main algorithm is at least $m - 5 \cstg \cdot m^{5/6}$, which is what we want. To see this claim, notice that in this situation there are two cases:
 	
 	\paragraph{Case 1:} There is an event $E_i$ ($i \le \bar{r}$) which does not hold because $\tilde{m}_i \le 5 \cstg \cdot  m^{5/6}$. But this means that at the beginning of round $i$ the main algorithm has already opened $m - \tilde{m}_i \ge m - 5 \cstg \cdot  m^{5/6}$ machines, and the claim holds.
 	
 	\paragraph{Case 2:} All the events $\{E_i\}_{i \le \bar{r}}$ do not hold because $\tilde{m}_{i+1} \le \tfrac{3}{4} \tilde{m}_i$ for all of them. But this means that in beginning of the last round $\bar{r}$ there are $$\tilde{m}_{\bar{r}} \,\le\, (\tfrac{3}{4})^{\bar{r}-1} \cdot \tilde{m}_1 \,=\, (\tfrac{3}{4})^{\bar{r}-1}\cdot m \,=\, \tfrac{4}{3} \,\le\, 5 \cstg \cdot m^{5/6}$$ unopened machines, and so the claim also holds. 
 	
 	So to prove \Cref{lemma:stochLogCalls} it suffices to show that the probability that an event $\{E_i\}_{i \le \bar{r}}$ holds is at most $O(\frac{1}{m})$. This probability is 
 	\begin{align}
 	\Pr\bigg(\bigvee_{i \le \bar{r}} E_i\bigg) \,\le\, \sum_{i \le \bar{r}} \Pr(E_i).  \label{eq:unionE}
 	\end{align}
 	To upper bound the right-hand side, we have $$\Pr(E_i) = \Pr(\tilde{m}_{i+1} > \tfrac{3}{4} \tilde{m}_i \mid \tilde{m}_i > 5 \cstg \cdot  m^{5/6}) \Pr(\tilde{m}_i > 5 \cstg \cdot  m^{5/6}).$$ But conditioning on $\tilde{m}_i > 5 \cstg \cdot  m^{5/6}$ (or more precisely, conditioning on the demands up to the beginning of round $i$ so that this event holds) and applying \Cref{lemma:openCst} (notice that even with this conditioning the items within round $i$ are still sampled i.i.d. from $\mu$) we have that the first term in the right-hand side is at most $\frac{\log(m/c(\mu))}{m^2}$; so this gives $\Pr(E_i) \le \frac{\log(m/c(\mu))}{m^2}$. Employing this on \eqref{eq:unionE} gives 
 	\begin{align*}
 	\Pr\bigg(\bigvee_{i \le \bar{r}} E_i\bigg) \,\le\, \bar{r} \cdot \frac{\log(m/c(\mu))}{m^2} \,=\, O\bigg(\frac{\log^2 m}{m^2} \bigg) \,=\, O\bigg(\frac{1}{m} \bigg).
 	\end{align*}
 	This concludes the proof of \Cref{lemma:stochLogCalls}.  
 \end{proof}

\section{Properties of \eqref{eq_lpconfigmin}}\label{sec_solve_lp}

 \subsection{Consistency of $\lpconfigmin$ with respect to  type partitions} \label{sec:consistencyLP}

In principle, the definition of \eqref{eq_lpconfigmin} depends on how the demands are partitioned into types. However, we show that this is not actually the case.  

\begin{lemma}\label{lem:types}
	The optimal value $\lpconfigmin$ is the same for every possible type partition of the demands.
\end{lemma}

In particular, the optimal value $\lpconfigmin$ is well-defined. To prove the lemma, it suffices to show that merging two demand types (that can be merged) does not change the optimal value.

	\begin{lemma}\label{lem:merge}
		Consider an instance $S = (s_1, \ldots, s_k)$ and a valid assignment of types $type : S \rightarrow \{0,1,\ldots,k\}$ (i.e. $s = s'$ whenever $type(s) = type(s')$) and such that types 0 and 1 have items of the same size (i.e. $s = s'$ whenever $type(s), type(s') \in \{0,1\}$). Let $\widetilde{type}$ be the type assignment that merges types 0 and 1, i.e. $\widetilde{type}(s) = 1$ for all $s$ such that $type^{-1}(s) \in \{0,1\}$ and $\widetilde{type}(s) = type(s)$ for the other $s$'s.
		
		Then the optimal value of $\lpconfigmin$ based on $type$ and $\widetilde{type}$ are the same. 
	\end{lemma}
	
	\begin{proof} \knote{To see if we want to make another pass, if anything is suspicious}
		Let $n_j = |type^{-1}(S)|$ be the number of items of type $j$ under the assignment $type$ and define $\tilde{n}_j$ analogously w.r.t. $\widetilde{type}$; notice $\tilde{n}_1 = n_0 + n_1$ and $\tilde{n}_j = n_j$ for all $j \ge 2$. Further, for any configuration $C$ w.r.t. $type$, we define $n_t(C)$ to be the number of demands of type $t$ in $C$. We define $\tilde{n}_t(\tilde{C})$ analogously for a configuration $\tilde{C}$ w.r.t $\widetilde{type}$. The LP values are then given by 
		\begin{multicols}{2}
			\begin{align*}
				LP := \min &\sum_C x_C\\
				st& \sum_C n_t(C) \cdot x_C \ge 2 n_t ~~~~~\forall t\\
				&x \ge 0
			\end{align*}    
			
			\begin{align*}
				\widetilde{LP} := \min &\sum_{\tilde{C}} x_{\tilde{C}}\\
				st& \sum_{\tilde{C}} \tilde{n}_t(\tilde{C}) \cdot x_{\tilde{C}} \ge 2 \tilde{n}_t~~~~~\forall t \geq 1\\
				&x \ge 0,
			\end{align*}    
		\end{multicols}
		where the configurations $C$ and $\tilde{C}$ are respectively in $\R^{k+1}$ and $\R^k$. 
		
		The second LP is a relaxation of the first, since it follows by adding the first two inequalities ($t=0$ and $t=1$) of the first LP. Thus, $\widetilde{LP} \le LP$. 
		
		Now we prove that $LP \le \widetilde{LP}$. Consider an optimal solution $\tilde{x}$ for the second LP. Given a valid configuration $\tilde{C}$ for the second LP, define the configurations $C^0 = (\tilde{C}_1, 0, \tilde{C}_2, \ldots,\tilde{C}_n)$ and $C^1 = (0, \tilde{C}_1, \tilde{C}_2, \ldots,\tilde{C}_n)$ that respectively assign all the items of $\widetilde{type}$ to $type$ 0 and 1. Consider the solution $x$ for the first LP given by $$x_{C^0} = \frac{n_0}{n_0 +n_1} \tilde{x}_{\tilde{C}},~~~~x_{C^1} = \frac{n_1}{n_0+n_1} \tilde{x}_{\tilde{C}}~~~~ \textrm{for all $\tilde{C}$}$$ and $x(C)=0$ for all other configurations. 
		
		We claim that $x$ is a feasible solution for the first LP with value $\widetilde{LP}$. For its value $$\sum_C x_C = \sum_{\tilde{C}} (x_{C^0} + x_{C^1}) = \sum_{\tilde{C}} \tilde{x}_{\tilde{C}} = \widetilde{LP},$$ as claimed. For its feasibility, for any $t \in \{0,\ldots,k\}$ we have
		\begin{align*}
			\sum_C n_t(C) \cdot x_C = \sum_{\tilde{C}} (n_t(C^0) \cdot x_{C^0} +n_t(C^1) \cdot x_{C^1}) = \sum_{\tilde{C}} \bigg(\frac{n_0}{n_0+n_1} n_t(C^0) + \frac{n_1}{n_0+n_1} n_t(C^1)\bigg) \cdot \tilde{x}_{\tilde{C}}.
		\end{align*}
		When $t \in \{0,1\}$, we see that $n_t(C^t) = \tilde{n}_1(\tilde{C})$ and the other term $n_t(C^{1-t})$ is zero. Hence 
		\begin{align*}
			\sum_C n_t(C) \cdot x_C = \frac{n_t}{n_0 + n_1} \sum_{\tilde{C}} \tilde{n}_1(\tilde{C}) \cdot \tilde{x}_{\tilde{C}} \ge \frac{n_t}{n_0+n_1} \cdot 2 \tilde{n}_1 = 2 n_t, 
		\end{align*}        
		where the first inequality is from the feasibility of $\tilde{x}$. So $x$ satisfies the constraints of the first LP when $t=0,1$. For the remaining constraints $t \ge 2$ we have that $n_t(C^0) = n_t(C^1) = \tilde{n}_t(\tilde{C})$, so 
		\begin{align*}
			\sum_C n_t(C) \cdot x_C =  \sum_{\tilde{C}} \tilde{n}_t(\tilde{C}) \cdot \tilde{x}_{\tilde{C}} \ge  2 \tilde{n}_t = 2 n_t, 
		\end{align*}        
		which are then satisfied as well. This proves that $x$ feasible for the first LP.
		
		We conclude, the optimal value of this LP is at most that of this solution $x$, which then gives $LP \le \widetilde{LP}$ as desired. 
	\end{proof}

\subsection{Solving \eqref{eq_lpconfigmin}}\label{sec:solve}

In this section, we show how to efficiently solve $\lpconfigmin$ up to small additive error. We need the following theorem of Rothvoss (stated in simplified form) \cite{rothvossConfigLP}, which relies on the Plotkin-Shmoys-Tardos algorithm to solve implicit fractional covering problems \cite{DBLP:journals/mor/PlotkinST95}.

\begin{thm}\label{thm_rothvoss_lp}
	Let $\mathcal{S} \subset 2^{[n]}$ be a set family. Suppose that we can solve the following \textsc{Subproblem}: Given parameter $\epsilon \in (0,1)$ and $y \in \mathbb{Q}^n_+$, output a set $S^* \in \mathcal{S}$ with $\sum_{i \in S^*} y_i \geq (1 - \epsilon) \cdot \max_{S \in \mathcal{S}} \big( \sum_{i \in S} y_i \big)$ in time $T(n, \epsilon)$.
	
	Then for any $\delta \in (0, n/2]$, we can find a basic solution of the following LP:
	    \begin{equation}\tag{$LP$}
		\begin{array}{rrll}
			\min & \sum_{S \in \mathcal{S}} x_S\\
			s.t. & \sum_{S \in \mathcal{S}} 1_{j \in S} \cdot x_S & \geq 1 &\quad \forall j \in [n]\\
			& x & \geq 0
		\end{array} 
	\end{equation}
	with objective value at most $LP + \delta$ in time $poly(n, 1/ \delta) \cdot T(n, \Omega(\delta / n))$.
\end{thm}

\textsc{Subproblem} is an approximate dual separation problem for $LP$. To apply this theorem, we relate \eqref{eq_lpconfigmin} to a ``per-demand'' configuration LP, which we can apply the theorem to.

\begin{lemma}\label{lem:solveapprox}
	Given \eqref{eq_lpconfigmin} defined on $n$ demands (with some partition into demands types) and a parameter $\delta \in (0, n/2]$, we can efficiently find a basic solution of \eqref{eq_lpconfigmin} with objective value at most $\lpconfigmin + \delta$ in time $poly(n, 1/\delta)$.
\end{lemma}
\begin{proof}
	We let $C$ be the collection of feasible  configurations with respect to that partition into demand types in \eqref{eq_lpconfigmin} (i.e. $C$ indexes all variables used by this LP.) We first define the natural ``per-demand'' configuration LP by taking each demand as its own type. Indexing the demands by $j \in [n]$, this LP is:
	\begin{equation}\tag{$LP_J$}
		\begin{array}{rrll}
			\min & \sum_{c' \in C_J} x_{c'}\\
			s.t. & \sum_{c' \in C_J} 1_{j \in c'} \cdot x_{c'} & \geq 2 &\quad \forall j \in [n]\\
			& x & \geq 0
		\end{array} 
	\end{equation}
	, where $C_J$ is the collection of all configurations where each demand is its own type. By \Cref{lem:types}, we have $\lpconfigmin = LP_J$. To apply the theorem, we re-scale the right hand side of $LP_J$ by dividing by $2$. Let $LP_J'$ be the resulting LP. Note that the extreme points and optimal solutions of $LP_J$ and $LP_J'$ are also related by a multiplicative $2$-factor.
	
	We now show how to solve \textsc{Subproblem} for $LP_{J}'$. We are given parameter $\epsilon \in (0,1)$ and $y \in \mathbb{Q}^n_+$, we must find a configuration in $C^* \in C_J$ with $\sum_{j \in C^*} y_j \geq (1 - \epsilon) \cdot \max_{C \in C_J} \big( \sum_{j \in C} y_j \big)$. Recall that the configurations in $C_j$ are exactly the subsets of demands $C \subset [n]$ with $\sum_{j \in C} s_j \leq 1$ and $\sum_{j \in C} s_j + \max_{j \in C} s_j \leq B$. Our algorithm for the subproblem is the following:
	
	\begin{enumerate}
		\item Guess the index $j^*$ of the largest-sized item used by the configuration achieving the maximum $\max_{C \in C_J} \big( \sum_{j \in C} y_j \big)$.
		\item Define the knapsack instance with demands of size at most $s_{j^*}$ in $[n] - \{j^*\}$ such that each remaining demand $j$ has size $s_j$ and value $y_j$. The knapsack size is $\min(1 - s_{j^*}, B - 2s_{j^*})$. Let $v^*$ be the optimal value of this knapsack instance. Run the knapsack FPTAS to obtain a subset of demands $\bar{C} \subset [n] - \{j^*\}$ with value at least $(1 - \epsilon) \cdot v^*$ in time $poly(n, 1/\epsilon)$.
		\item Output the demands $\bar{C} \cup \{j^*\}$.
	\end{enumerate}
	There are $n$ guesses for $j^*$, so the algorithm runs in time $poly(n, 1 / \epsilon)$. For correct guess of $j^*$, the algorithm outputs $\bar{C} \cup \{j^*\}$, which is a feasible configuration by definition of the residual knapsack instance (we only use demands of size at most $s_{j^*}$ =, so $max_{j \in \bar{C} \cup \{j^*\}} s_j = s_{j^*}$ and the knapsack budget ensures the required constraints.) Further, we have $\sum_{j \in \bar{C} \cup \{j^*\}} \geq (1 - \epsilon) \cdot v^* + y_{j^*} \geq (1 - \epsilon) \cdot \max_{C \in C_J} \big( \sum_{j \in C} y_j \big)$. Thus, we can solve \textsc{Subproblem} for $LP_J'$ in time $T(n, \epsilon) = poly(n, 1/ \epsilon)$.
	
	Now we can apply the theorem to $LP_J'$ to obtain an extreme point of $LP_J'$, say $x^*$, with $\sum_{c' \in C_J} x^* \leq LP_J' + \delta$ in time $poly(n, 1/ \delta)$. It follows, $2x^*$ is an extreme point of $LP_J$ with objective value at most $2 \cdot LP_J' + \delta = LP_J + \delta$.
	
	Because $2x^*$ is an extreme point of $LP_J$, which has $n$ non-trivial constraints, $2x^*$ has at most $n$ non-zero variables. Let $\bar{C}_J = \{c' \in C_J \mid 2x^*_{c'} > 0\}$ be the sub-collection of configurations used by this extreme point. Then $\lvert \bar{C}_J \rvert \leq n$. Now, consider modifying $LP_J$ by keeping only the variables indexed by $\bar{C}_J$. Let the resulting LP be $LP_J(\bar{C}_J)$. We have $LP_J(\bar{C}) \leq LP_J + \delta$ because they share the solution $2x^*$.
	
	Finally, we relate $LP_J(\bar{C}_J)$ with $\lpconfigmin$. First, we map the ``per-demand'' configurations of $\bar{C}_J$ to the ``per-type'' configurations of $C$ as follows: Suppose the demands $J$ are partitioned into types $J = \cup_t J_t$. Then For each configuration $c' \in \bar{C}_J$, we map $c'$ to a configuration in $C$ with $\lvert c' \cap J_t \rvert$-many demands of type $t$ for every type $t$. Note that this can be done efficiently because we $\lvert \bar{C}_J \rvert \leq n$. By definition, the mapped configuration is feasible. Let $\bar{C} \subset C$ be the all per-type configurations that are mapped to by some configuration in $\bar{C}_J$. Then $\lvert \bar{C} \rvert \leq n$. Further, we have $\lpconfigmin(\bar{C}) = LP_J(\bar{C}_J)$ by an analogous argument as in \Cref{lem:types}.
	
	To conclude, let $\lpconfigmin(\bar{C})$ be obtained from $\lpconfigmin$ by keeping only variables indexed by $\bar{C}$. Then, $\lpconfigmin(\bar{C})$ has polynomially many variables and constraints, so we can explicitly solve $\lpconfigmin(\bar{C})$ to obtain an optimal extreme point, which is also an extreme point of $\lpconfigmin$ with objective value $\lpconfigmin(\bar{C}) = LP_J(\bar{C}_J) \leq LP_J + \delta = \lpconfigmin + \delta$, as required.
\end{proof}

\end{document}